\documentclass[a4paper,12pt,twoside]{article}
\usepackage{epsfig,graphicx,color}
\usepackage{cite}
\usepackage{floatflt,indentfirst}
\usepackage{amsfonts,amssymb,latexsym,amsmath,enumerate,amsthm}
\usepackage[mathscr]{eucal}
\def\thesection{\arabic{section}.}
\def\theequation{\thesection\arabic{equation}}

\makeatletter \@addtoreset{equation}{section}
\makeatother \allowdisplaybreaks

\begin{document}
\def\rmd{\mathrm d}
\def\rme{\mathrm e}

\title{\large\bf
Pattern formation in terms of semiclassically limited distribution
on lower-dimensional manifolds for nonlocal
Fisher--Kolmogorov--Petrovskii--Piskunov equation}

\author{Levchenko$^{1}$ E.A.,
Shapovalov$^{2}$ A.V.,  and Trifonov$^{1}$ A.Yu.}

\date{\normalsize$^1$ Laboratory of Mathematical Physics\\  of Mathematical Physics Department,\\
Tomsk Polytechnical University,\\
Lenin ave. 30, Tomsk, 634050, Russia\\\vskip 0.4cm
$^2$ Theoretical Physics Department,\\
Tomsk State University,\\
Lenin ave. 36, Tomsk, 634050, Russia\\\vskip 0.4cm
{\small e-mail: levchenkoea@tpu.ru, e-mail: shpv@phys.tsu.ru,
e-mail: atrifonov@tpu.ru}}
\maketitle

\begin{abstract}

We have investigated the pattern formation in systems described by
the nonlocal Fisher--Kolmogorov--Petrovskii--Piskunov equation for
the cases where the dimension of the pattern concentration area is
less than that of independent variables space. We have
obtained a system of integro-differential equations which describe
the dynamics of the concentration area and the semiclassically
limited distribution of a pattern in the class of trajectory
concentrated functions. Also, asymptotic large-time solutions have
been obtained that describe the semiclassically limited distribution
of a quasi-steady-state pattern on the concentration manifold. The
approach is illustrated by an example for which the analytical
solution is in good agreement with the prediction of a numerical
simulation.

\end{abstract}

\section*{Keywords}
Pattern formation, nonlocal population dynamics, Fisher--Kolmogorov--Petrovskii--Piskunov
equation, semiclassical approximation.

\section*{Introduction}

Nonlocal reaction-diffusion (RD) models are generally used to
describe structures ordered in space and time. Structures of this
type, formed by self-organization mechanisms, are involved in many
important phenomena in biology, medicine, epidemiology, and ecology,
such as the pattern formation in population dynamics, cancer
treatment, evolution of infectious diseases, etc. (see, e.g., the
review papers \cite{lee,math-for-life}, and references therein).

Evolution of one-species microbial populations with long-range
interactions between individuals  is modeled by nonlocal
generalizations of the classical
Fisher--Kolmogorov--Petrovskii--Piskunov (FKPP) equation
\cite{fisher1937,kolmogorov1937} in the population density $u(x,t)$:
\begin{equation}
u_t(x,t)=D\Delta u(x,t)+au(x,t)- bu^2(x,t). \label{intr1}
\end{equation}
Equation (\ref{intr1}) contains the terms that describe diffusion
with coefficient $D$, population growth with rate $a$ and local
competition with rate $b$.

Nonlocal effects arise in competitive interactions of microbial
populations due to the diffusion of nutrients, the release of toxic
substances, chemotaxis, and molecular communications among
individuals
\cite{lee,math-for-life,fuentes2003,fuentes2004,kenkre2004}.

No space ordered structures (patterns) occur during the evolution of
the system governed by equation (\ref{intr1}). In the nonlocal FKPP
models, patterns appear due to nonlocal competitive losses and
diffusion \cite{fuentes2003, fuentes2004, kenkre2004}, convection
\cite{cunha2009} and nonlocal growth \cite{cunha2011} under certain
choice of parameters. Note that the pattern formation in the
nonlocal FKPP models is different from the well-known Turing
morphogenesis where the mechanism of pattern formation relies on the
competition between the activator and the inhibitor
\cite{turing,Murray2001}.

In this work we consider the following version of the nonlocal FKPP equation:
\begin{multline}
u_{t}=D\Delta u-\bigg\langle\nabla ,u[V_{\vec x}(\vec x,t)+\varkappa
\int\limits_{{\mathbb R}^n}W_{\vec x}(\vec x, \vec y, t)u(\vec y,
t){\rmd}\vec y]\bigg \rangle+\\
 +a(\vec x,t)u-\varkappa u \int\limits_{{\mathbb
R}^n}b_\gamma(\vec x, \vec y)u(\vec y, t){\rmd}\vec y, \label{eq1}
\end{multline}
where $u(\vec x,t)$ is a smooth scalar function belonging to a
Schwartz space $\mathbb S$ in the space variable $\vec x\in{\mathbb R}^n$
at each point in time $t$, $\langle\vec a, \vec b \rangle$ is the
Euclidian scalar product of $\vec a,\vec b \in{\mathbb R}^n$,
$|\vec a|^2=\langle\vec a, \vec a \rangle$. Here, the local
competition term in \eqref{eq1} has been replaced by the the term of
nonlocal losses controlled by the influence function
$b_\gamma(\vec x,\vec y)$ with a range parameter $\gamma$.

External factors can cause convective processes, which contribute to
the population dynamics \cite{cunha2009,cunha2011}. The
gradient vectors $V_{\vec x} = \nabla_x V(\vec x,t)$ and
$W_{\vec x} = \nabla_xW(\vec x,\vec y,t)$ in equation \eqref{eq1} describe the local and the
nonlocal convective forces, respectively. In a bacteria population,
the nonlocal convection term describes the flow of bacteria that
move under the action of the force given by the gradient of the
potential $W$ produced by other bacteria \cite{clerc2006}.

The nonlocal FKPP equations were treated analytically and
numerically by several authors.

Equation \eqref{eq1} was solved numerically for Gaussian and cutoff
influence functions $b_\gamma(\vec x,\vec y)$ with periodic boundary
conditions in a 2D case and with a null flow boundary condition on
the in a 1D case \cite{fuentes2003}. Spatial structures were
obtained and analyzed for a given relationship between the width of
the influence function and the size of the population domain.

The stability of homogeneous steady-state 1D solutions was examined
\cite{fuentes2004} by using dispersion relation between the
wavenumber of any mode of the pattern and the rate of its grow.

The transition from a homogeneous steady-state to a spatially
modulated stable state was considered in \cite{maruvka2006}.
The spatial invasion of a stable into an unstable phase was studied
for a branching-coalescence process with nonlocal competition
\cite{maruvka2007}.

The nonlocal FKPP equation (\ref{eq1}) with the cutoff influence
function $b_\gamma(\vec x,\vec y)$ and the convection caused by constant
and spatial velocity fields was investigated for a case where
diffusion was not significant for the pattern formation
\cite{cunha2009}.

Limit values of the parameters for the cases of patterns appearing
in the presence of convection were estimated using the dispersion
relation obtained by the perturbation method for the 1D equation
(\ref{eq1}) similar to that used in \cite{fuentes2004}. The
influence of convection on the pattern formation was investigated
numerically.

In \cite{genieys2007} the nonlocal 1D FKPP equation is used for
studying of pattern formation in the problem of ecological invasion
where space variable $x$ is treated as a physiological trait.

Nonlocal interactions in one-species RD systems can also manifest
themselves as population traveling waves \cite{volpert2009, mei2011,
maruvka2006, maruvka2007}, swarm formation \cite{billingham2004},
etc.

From the above references we see that only some properties of the
pattern formation can be investigated analytically, such as the
necessary conditions of their emergence and sustainability issues.
General view of the whole structure provides  a numerical 1D
solution of FKPP equation. The solution construction and analysis of
the pattern properties depending on the model parameters becomes
much more complicated in the multidimensional case.

In this paper, we investigate patterns described by
equation (\ref{eq1}) and concentrated on manifolds dimension of
which $k$ is less than the number of independent variables in the
equation $n$, $k<n$. Such patterns can be studied using a system
of equations describing evolution of the pattern concentration area.

For easy consideration, we restrict ourself to a simply connected manifold
\begin{equation}
\Lambda^k_t=\bigg(\vec x \in {\mathbb R}^n\bigg|\vec x=\vec X(t,s),
s\in {\mathbb G}\subset{\mathbb R}^k \bigg). \label{mnog}
\end{equation}
Here, the real variables $s$, $s\in{\mathbb G}\subset{\mathbb R}^k$,
parametrize the manifold $\Lambda^k_t$, the real vector $\vec
X(t,s)$ smoothly depends on $t\in {\mathbb R}^1$ and
parameters $s$.

The manifold  $\Lambda^k_t$ carries information about the evolution
of the pattern geometry. A similar approach is embodied in the
Cartan's method of moving frames in which a moving frame is adapted
to the kinematic properties of the observer in motion.

Manifolds naturally arise as concentration domains for the solutions
of multidimensional (integro-) differential equations in the
WKB-Maslov formalism of semiclassical asymptotics
\cite{Maslov1,Maslov2,BelDob}. Lower dimensional manifolds with a
complex germ \cite{Maslov2,BelDob} allow one to construct asymptotic
solutions of the original equation (\ref{eq1}) for $D \to 0$. This
approximation of small diffusion seems to be quite reasonable (see,
e.g. \cite{cunha2009} where diffusion is neglected at all).

The solution $u(\vec x,t)$ of equation (\ref{eq1}) generates on the
manifold $\Lambda^k_t$ a distribution $\rho(t,s)$, which can be
assumed to be a \emph{semiclassically limited distribution} (SLD), as $D
\to 0$, in the space ${\mathbb{R}}^k$. The SLD is determined by
simpler equations compared to the original equation (\ref{eq1}), and
it carries the most significant information about the pattern.

In terms of the semiclassical formalism, we consider here a special
case of the 2D pattern formation.
For this purpose we consider the 2D equation (\ref{eq1}) in a class
of functions concentrated in a neighborhood of a 1D curve in a 2D space
(${\mathbb{R}}^2$).

In Section 1, we describe a lower-dimensional manifold $\Lambda_t^k$
where solutions of the nonlocal FKPP equation (\ref{eq1}) are concentrated.
In Section 2, a dynamic system describing evolution of $\Lambda_t^k$ and
$\rho(t,s)$ is deduced. In Section 3, we propose a method of solution
of the dynamic system for $V(\vec x, t)=W(\vec x, \vec y, t)=0$ in
(\ref{eq1}). In Section 4, we construct an exact
solution of the dynamic system with the symmetric influence function
$b_\gamma(\vec x,\vec y)$ in the 2D case. In Section 5, a class of asymptotic solutions
is found. These solutions are perturbations of the exact solution of Section 4 and
tend to this solution as $T \to \infty$. Some of the asymptotic solutions are treated
as pattern formation description. In Section 6, we
consider evolution of the SLD $\rho(t,s)$ with diffusion. In Conclusion,
basic results are discussed.

\section[Concentration manifold]{Concentration manifold}

Geometric properties of bacterial patterns \cite{Murray2001} are
characterized by the bacterial density distribution on a geometric
object (manifold). For instance, a bacterial colony having a
density maximum at a point is concentrated in à neighborhood of the
point (zero-dimensional manifold), a ring distribution (see, e.g.,
\cite{budrene1988,matsushita2004,cyganov2001}) is concentrated in à
neighborhood of a circumference (one-dimensional manifold), etc.
Generally, the concentration manifold of a pattern and the SLD on the
manifold entire important characteristics of the pattern that
carry information about the entire density distribution.

From a mathematical point of view, a pattern is described by the
solution of equation (\ref{eq1}), and so it is necessary to describe
concentration domain of the solution and find the SLD.

Define a class $J_D(\Lambda^k_t)$ of functions $u(\vec x, t, D)\in
J_D(\Lambda^k_t)$ depending on parameter $D$ and concentrated on a
manifold $\Lambda^k_t$. Suppose that functions $u(\vec x, t, D)$
decrease as $|\vec x|\to \infty$ faster than any power of $\vec x$,
and so moments of any finite order exist for these functions.

For any smooth function $A(\vec x,t)$ and $u(\vec x, t, D)\in
J_D(\Lambda^k_t)$, we define
\begin{equation}
A_u(t,D)=\dfrac{1}{m_u(t,D)}\int\limits_{{\mathbb R}^n}A(\vec x,
t)u(\vec x, t, D){\rmd}\vec x , \label{eq3}
\end{equation}
where $m_u(t,D)$ is the zero moment of the function $u(\vec x, t, D)$:
\begin{equation}
 m_u(t,D)=\int\limits_{{\mathbb R}^n} u(\vec x,t,D){\rmd}\vec x. \label{eqmu}
\end{equation}
Assume that there exist a limit
\begin{equation}
\lim_{D\to 0} m_u(t,D)=\lim_{D\to 0}\int\limits_{{\mathbb R}^n}
u(\vec x,t,D){\rmd}\vec x=\int\limits_{{\mathbb G}} \rho(t,s){\rmd}s
\label{eqmu-1}
\end{equation}
and denote
\begin{equation}
m_\rho(t)=\int\limits_{{\mathbb G}} \rho(t,s){\rmd}s.
\label{eqmu-1-1}
\end{equation}

As $u(\vec x,t,D)$ has the meaning of population density, it follows
that $u(\vec x,t,D)$ and $\rho(s,t)$ are non-negative.

We say that a function $u(\vec x,t,D)$ belongs to the class
$J_D(\Lambda^k_t)$ if
\begin{equation}
\lim_{D\to 0}A_u(t,D)=\dfrac{1}{m_{\rho}(t)}\int\limits_{\mathbb
G}A(\vec X(t,s),t)\rho(t,s){\rmd}s. \label{eq2}
\end{equation}

Following \cite{bagre}, we refer to $J_D(\Lambda^k_t)$ as the class
of functions semiclassically concentrated on a manifold
$\Lambda^k_t$. The solutions of equation (\ref{eq1}) found in this
class describe the patterns above, i.å. patterns concentrated in the
neighborhood of the manifold $\Lambda^k_t$.

Equations (\ref{eq2}) and (\ref{eq3}) can be written in the
equivalent form
\begin{equation}
\lim_{D\to 0}\dfrac{u(\vec x, t, D)}{m_u(t,D)}=\int\limits_{\mathbb G}
\delta(\vec x-\vec X(t,s))\rho(t,s){\rmd}s. \label{eq31a}
\end{equation}

Note that the functions $\rho(t,s)$ and $u(\vec x,t,D)$ are
explicitly connected. Let $(s,\xi)$ be a coordinate system in the
space ${\mathbb R}^n$ where the variables $\xi\in {\mathbb
W}\subset{\mathbb R}^{n-k}$ complement the variables $s$ to the
coordinate system in ${\mathbb R}^n$, i.e. $\vec x=\vec x(s,\xi)$,
$s=s(\vec x)$, $\xi=\xi(\vec x)$, and Jacobian
$J\Bigl(\dfrac{\vec x(s,\xi)}{s,\xi}\Bigr)\ne 0$. Define the variables
$\xi$ so that their coordinate lines are orthogonal to the manifold
$\Lambda^k_t$ with respect to the Euclidean inner product in a
tangent space. So we obtain
\begin{equation}
\rho(s,t)=\lim_{D\to 0}\int\limits_{\mathbb W} u(\vec x(s,\xi),t,D)
J\bigg(\dfrac{\vec x(s,\xi)}{s,\xi}\bigg){\rmd}\xi. \label{eq3a}
\end{equation}

\section[Evolution of the manifold]{Evolution of the manifold}

Let us obtain a system of equations to describe the evolution of a
function $\rho(t,s)$ and a vector $\vec X(t,s)$ related to a
solution $u(\vec x,t, D)$ of equation (\ref{eq1}) in the class
$J_D(\vec X(t,s))$.

According to (\ref{eq3}), we define the first normalized moment of
the function $u(\vec x,t,D)$ as
\begin{equation}
\vec x_u(t,D)=\dfrac{1}{m_u(t,D)}\int\limits_{{\mathbb R}^n} \vec x u(\vec x,t,D){\rmd}\vec x. \label{FPK4'}
\end{equation}

Using (\ref{eq2}) and (\ref{FPK4'}), we obtain
\begin{equation}
\vec x_{\rho}(t)= \lim_{D\to 0} \vec x_u(t,D)=\dfrac 1 {m_{\rho}(t)}
\int\limits_{\mathbb G} \vec X(t,s)\rho (t,s){\rmd}s. \label{evol2}
\end{equation}

From (\ref{eqmu-1}) and (\ref{evol2}) it follows that
\begin{align}
& \dot {m}_\rho=\int\limits_{\mathbb G} \dot\rho(t,s){\rmd}s, \label{mklassik0}\\
& \dot{\vec x}_\rho= -\dfrac{\dot
{m}_\rho}{m_\rho}\vec x_\rho(t,D)+ \dfrac{1}{m_\rho} \int \limits_{\mathbb G}
\big[(\dot {\vec X}(t,s)\rho (t,s) + \vec X(t,s)\dot \rho (t,s)\big]
{\rmd}s. \label{xclassic0}
\end{align}

On the other hand, differentiating equations (\ref{eqmu}) and
(\ref{FPK4'}) with respect to $t$ and taking into account
(\ref{eq1}), we obtain
\begin{align}
& \dot {m}_u=\int\limits_{\mathbb{R}}u_t(\vec x,t) d\vec x=
\int\limits_{\mathbb{R}}\Big[ a(\vec x,
t)-\varkappa
\int\limits_{\mathbb{R}} b_\gamma(\vec x, \vec y)u(\vec y, t,D)d\vec y \Big]u(\vec x, t,D) d\vec x, \label{mklassik}\\
& \dot{\vec x}_u= \dfrac{1}{m_u}\int\limits_{{\mathbb R}^n}\Big(
V_{\vec x}(\vec x,t)+\varkappa\int\limits_{{\mathbb R}^n}W_{\vec x}(\vec x,\vec y,t)
u(\vec y,t,D)d\vec y \Big)u(\vec x,t,D)d\vec x + \nonumber\\
& +\dfrac{1}{m_u}\int\limits_{{\mathbb R}^n} (\vec x-\vec x_u)\Big[ a(\vec x,
t)-\varkappa\int\limits_{\mathbb{R}} b_\gamma(\vec x, \vec y)u(\vec y, t)d\vec y \Big]u(\vec x,t,D) d\vec x. \label{xclassic}
\end{align}

The equations describing the evolution of the SLD $\rho(t,s)$ and of
the vector $\vec X(t,s)$ are obtained by a limiting process at
${D\to 0}$ in equations (\ref{mklassik}) è (\ref{xclassic}),
respectively:
\begin{align}
& \dot \rho (t,s)=\rho(t,s)\big[a(\vec X(t,s),t)-\varkappa
\int\limits_{\mathbb G}b_\gamma(\vec X(t,s),\vec X(t,s'))\rho(t,s')ds'\big] , \label{evolmn1}\\
& \dot {\vec X} (t,s)=V_{\vec x}(\vec X(t,s),t)+\varkappa\int\limits_{\mathbb
G} W_{\vec x}(\vec X(t,s),\vec X(t,s'),t)\rho(t,s')ds'. \label{evolmn2}
\end{align}

From (\ref{evolmn2}) it follows that dynamics of the manifold
$\Lambda^k_t$ is determined by the convective terms $V_{\vec x}(\vec
x,t)$ and $W_{\vec x}(\vec x,\vec y,t)$ in the FKPP equation (\ref{eq1}).

The system of equations (\ref{evolmn1}), (\ref{evolmn2}) is closed
and describes evolution of the vector $\vec X(t,s)$ determining
the manifold $\Lambda^k_t$ and the SLD $\rho(t,s)$ on the manifold
$\Lambda^k_t$. To each solution $u(\vec x,t, D)$ of equation (\ref{eq1})
with an initial condition
\begin{equation}
u(\vec x,t, D)|_{t=0}=\varphi(\vec x,D)\label{evolmn3}
\end{equation}
there corresponds a solution of system (\ref{evolmn1}),
(\ref{evolmn2}) with the initial conditions
\begin{equation}
\rho(t,s)|_{t=0}=\rho_{\varphi}(s), \quad \vec X(t,s)|_{t=0}=\vec
X_{\varphi}(s), \label{evolmn4}
\end{equation}
where $\rho_{\varphi}(s)$ and $\vec X_{\varphi}(s)$ are related to
$\varphi(\vec x,D)$ by \eqref{eqmu-1} and \eqref{evol2},
respectively:
\begin{equation}
\lim_{D\to 0} m_\varphi(D)=m_{\rho_\varphi}=\int\limits_{\mathbb G}
\rho_\varphi(s,t){\rmd}s, \label{eqmu-2}
\end{equation}
\begin{equation}
\lim_{D\to 0} \dfrac{1}{m_\varphi(D)} \int\limits_{{\mathbb R}^n}
\vec x\varphi(\vec x,D){\rmd}\vec x=\dfrac1{m_{\rho_\varphi}}
\int\limits_{\mathbb G} \vec X_\varphi(s)\rho_\varphi(s){\rmd}s,
\label{evol-3}
\end{equation}
where
\begin{equation}
m_\varphi(D)=\int\limits_{{\mathbb R}^n}\varphi(\vec x,D){\rmd}\vec x.
\label{eqmu-2-1}
\end{equation}

We refer to equations (\ref{evolmn1}), (\ref{evolmn2}) as {\it the
Einstein–-Ehrenfest} (EE) dynamical system of $(k,M)$ type for
$M=1$. Here, $k$ is the dimension of the manifold $\Lambda^k_t$, $M$
is the highest order of the moments in the system.

Therefore, the study of patterns described by equation (\ref{eq1}) in
terms of the SLD $\rho(t,s)$ on the manifold $\Lambda^k_t$ is
reduced to solving the EE system (\ref{evolmn1}),
(\ref{evolmn2}) with initial conditions (\ref{evolmn4}).

\section[Solution of the Einstein--Ehrenfest system without convection]
{Solution of the Einstein--Ehrenfest system without convection}

Consider a method of solution of the Cauchy problem to the Einstein-Ehrenfest system
(\ref{evolmn1}), (\ref{evolmn2}) without convection, i.e. when
\begin{equation}
V(\vec x,t)=0, \quad W(\vec x,\vec y,t)=0,\quad \vec x\in{\mathbb{R}}^n
\label{evolmn-01}
\end{equation}
in equation \eqref{eq1}.

Then equation \eqref{eq1} takes the form
\begin{equation}
u_{t}=D\Delta u +a(\vec x,t)u-\varkappa u \int\limits_{{\mathbb
R}^n}b_\gamma(\vec x,\vec y)u(\vec y, t){\rmd}\vec y \label{eq1_1}
\end{equation}
and the EE system (\ref{evolmn1}), (\ref{evolmn2}) is
greatly simplified. From \eqref{evolmn2} it follows that
\begin{equation}
\vec X(t,s)=\vec X_{\varphi}(s). \label{evolmn1d}
\end{equation}
Substituting \eqref{evolmn1d} in equation \eqref{evolmn1} we get
\begin{align}
&\dot \rho (t,s)=\rho(t,s)\big[\tilde a(t,s)-\varkappa
\int\limits_{\mathbb G}
\tilde b_\gamma(s,s')\rho(t,s')ds'\big] ,\label{evolmn-dd}\\
&\rho(t,s)|_{t=0}=\rho_{\varphi}(s), \label{evolmn-ddn}
\end{align}
where
\begin{equation}
\tilde b_\gamma(s,s')=b_\gamma(\vec X_{\varphi}(s),\vec
X_{\varphi}(s')), \quad \tilde a(t,s)=a(\vec X_{\varphi}(s),t).
\label{evolmn-02}
\end{equation}

Consider an auxiliary linear problem of finding the eigenfunctions
$v_j(s)$ and eigenvalues $\lambda_j$ of a Fredholm equation with a kernel $\tilde
b_\gamma(s,s')$ (see, e.g., \cite{Vladimirov}):
\begin{equation}
\int\limits_{\mathbb G}\tilde b_\gamma(s,s')v_j(s')ds' = \lambda_j v_j(s).
\label{anal_resh2d}
\end{equation}
Here $j= (j_1,j_2,\dots,j_k)$ is multiindex, $j\in {\mathbb Z}^k_+$.

To be definite, we assume that \eqref{anal_resh2d} is the Fredholm
equation with à symmetrical kernel $\tilde b_\gamma(s,s')=\tilde
b_\gamma(s',s)$ and its solutions form an orthogonal system
\begin{equation}
\int\limits_{\mathbb G} v_l^*(s)v_k(s)ds=\delta_{lk}, \label{anal_ortog}
\end{equation}
where $v_l^*(s)$ is the complex conjugate of $v_l(s)$.
So, we can find solutions to equation \eqref{evolmn-dd} as an expansion
of $\rho(t,s)$ in terms of the eigenfunctions of the kernel $\tilde
b_\gamma(s,s')$ given by \eqref{evolmn-02}:
\begin{equation}
\rho(t,s)=\sum_{|j|=0}^{\infty}\beta_j(t)v_j(s).
\label{anal_resh3d}
\end{equation}
Similarly,
\begin{equation}
\rho_\varphi(s)=\sum_{|j|=0}^{\infty}\beta_{0j}v_j(s).
\label{anal_resh3b_1}
\end{equation}

In view of \eqref{anal_ortog}, the Fourier coefficients $\beta_j(t)$ and $\beta_{0j}$
are calculated as
\begin{eqnarray}
&& \beta_j(t)=\int\limits_{\mathbb G} v_j^*(s)\rho(t,s){\rmd}s, \label{anal_resh4d} \\
&& \beta_j(t)\big|_{t=0}=\beta_{0j} = \int\limits_{\mathbb G}
v_j^*(s)\rho_\varphi(s){\rmd}s. \label{anal_resh4dd}
\end{eqnarray}

The kernel $\tilde b_\gamma(s,s')$ of the form
\eqref{evolmn-02} can be represented as \cite{Vladimirov}
\begin{equation}
\tilde b_\gamma(s,s')=\sum_{|j|=0}^\infty
\lambda_jv_j(s)v_j^*(s').\label{anal_kernel}
\end{equation}
In view of \eqref{anal_resh2d}, \eqref{anal_resh3d} and \eqref{anal_resh4d},
equation \eqref{evolmn-dd} takes the form
\begin{equation}
\dot\rho
(t,s)=\rho(t,s)\Big[\sum_{|j|=0}^{\infty}\big[a_j(t)-\varkappa\lambda_j
\beta_j(t)\big]v_j(s)\Big]. \label{evolmn-ddd}
\end{equation}
Here we have used the notation
$$
a_j(t)=\int\limits_{\mathbb{G}} \tilde
a(t,s)v_j^*(s)ds.
$$

Differentiation of equation \eqref{anal_resh4d} with respect to $t$ yields
\begin{align*}
\dot\beta_j(t)&=
\int\limits_{\mathbb G}\dot\rho(t,s) v_j^*(s){\rmd}s=\\
&=\sum_{|j'|=0}^{\infty}\int\limits_{\mathbb
G}\rho(t,s)\Big(a_{j'}(t)-\varkappa\lambda_{j'}
\beta_{j'}(t)\Big)v_{j'}(s) v_j^*(s){\rmd}s.
\end{align*}

In view of the expansion
\begin{equation}
v_j^*(s)v_{j'}(s)=\sum_{|j''|=0}^{\infty}\Omega_{jj'}^{j''}
v_{j''}^*(s),\label{evolmn-03}
\end{equation}
we obtain system
\begin{equation}
\dot\beta_j = \sum_{|j'|=0}^{\infty}
[a_{j'}(t)-\varkappa\lambda_{j'}
\beta_{j'}]\sum_{|j''|=0}^{\infty}\Omega_{jj'}^{j''}\beta_{j''}.
\label{anal_resh4dd1}
\end{equation}
with the initial condition \eqref{anal_resh4dd}.

System \eqref{anal_resh4dd1} is equivalent to equation
\eqref{evolmn-dd}, and its solution can be found independently.
This property provides the way of solution of equation \eqref{evolmn-ddd}.
Namely, let solution of the Cauchy problem \eqref{anal_resh4dd1},
\eqref{anal_resh4dd} is known. Then solution of equation \eqref{evolmn-dd}
with initial condition \eqref{evolmn-ddn} is given
by \eqref{anal_resh3d}. In addition, we can obtain another
representation for the solution of the problem \eqref{evolmn-dd},
\eqref{evolmn-ddn} as:
\begin{equation}
\rho(t,s)=\rho_\varphi(s)\exp\biggl[\sum_{|j|=0}^{\infty}\int\limits_0^t
\big(a_j(\tau)-\varkappa\lambda_j
\beta_j(\tau)\big)v_j(s){\rmd}\tau\biggr].\label{evolmn-dddd}
\end{equation}

We next consider an example which illustrates the method described.

\section[Exact solution of the Einstein-Ehrenfest system]
{Exact solution of the Einstein-Ehrenfest system}

Let us construct an exact solution of equation \eqref{evolmn-dd} with coefficients
\eqref{evolmn-02} where
\begin{equation}
a(\vec x,t)=a={\rm const}, \quad b_\gamma(\vec x,\vec y)=b_0\exp
\Bigl\{-\dfrac{(\vec x-\vec y)^2}{2\gamma^2}\Bigr\}, \quad \vec x\in {\mathbb R}^2.
\label{anal1_0}
\end{equation}
The solution constructed is concentrated on the manifold $\Lambda_t^1$
determined by \eqref{evolmn1d} where
\begin{equation}
{\vec X}(t,s) = \vec X_\varphi(s)=(R \cos s,R \sin s), \quad s\in {\mathbb G} =[-\pi,\pi]\subset{\mathbb R}^1. \label{anal1}
\end{equation}

In this case, the manifold $\Lambda_t^1$ is compact, and equation
(\ref{evolmn-dd}) becomes
\begin{equation}
\dot\rho(t,s)=a\rho(t,s)-\varkappa\rho(t,s)\int\limits_{-\pi}^{\pi} \tilde b_\gamma(s,s')\rho(t,s'){\rmd}s', \label{anal2}
\end{equation}
where
\begin{equation}
\tilde b_\gamma(s,s')= b_0\exp \Bigl\{-\dfrac{(\vec X_\varphi(s)-\vec X_\varphi(s'))^2}{2\gamma^2}\Bigr\}
= b_0\exp\Bigl(-\dfrac{R^2}{\gamma^2}\big[1-\cos(s-s')\big]\Bigr).
\label{anal4a}
\end{equation}

The eigenfunctions $v_j(s)$ and eigenvalues $\lambda_j$ of the
Fredholm operator (\ref{anal_resh2d}) with the kernel $\tilde
b_\gamma(s,s')$ \eqref{anal4a} have the form \cite{Vladimirov}
\begin{equation}
v_j(s)=\dfrac{1}{\sqrt{2\pi}}e^{ijs}, \quad \lambda_j = 2\pi
b_0e^{-\mu}I_j(\mu), \quad j=\overline{-\infty,\infty},
\label{anal_resh3}
\end{equation}
where $\mu = R^2/\gamma^2$ and $I_j(\mu)$ is a modified Bessel
function of the first kind \cite{Beitman}. The functions $v_j(s)$
form an orthogonal system \eqref{anal_ortog}. Then the kernel
$\tilde b_\gamma(s,s')$ \eqref{anal4a} according to
\eqref{anal_kernel} can be written as
\begin{equation}
\tilde b_\gamma(s,s')=\sum_{j=-\infty}^{\infty}2\pi
b_0e^{-\mu}I_j(\mu)v_j(s)v_{-j}(s'). \label{anal_resh3a}
\end{equation}
Here, we take into account that $v_{j}^*(s)=v_{-j}(s)$.
In view of \eqref{anal_resh3} and \eqref{anal_resh3a}, equation
\eqref{evolmn-ddd} can be written as
\begin{equation}
\dot\rho(t,s)=a\rho(t,s)-\varkappa\rho(t,s)\bigg\{\sum_{j=-\infty}^{\infty}
\lambda_j\beta_{j}(t)v_j(s)\bigg\},\label{anal4x}
\end{equation}
where $\beta_j(t)$ is defined by (\ref{anal_resh4d}). From
(\ref{anal_resh3}) and (\ref{evolmn-03}) we get that
$\Omega_{jj'}^{j''}=(\sqrt{2\pi})^{-1}\delta_{j^{''}(j-j')}$
and equation (\ref{evolmn-03}) takes the form
\begin{equation}
v_{j}^*(s) v_j'(s) = \dfrac{1}{\sqrt{2\pi}}v_{-j+j'}(s).
\label{soot}
\end{equation}
Equations \eqref{soot} and (\ref{anal_resh4dd1}) yield the system
\begin{align}
\dot \beta_j =a \beta_j-\dfrac{\varkappa}{\sqrt{2\pi}} \sum_{l=-\infty}^{\infty}\lambda_l \beta_{j-l}\beta_{l}
,\quad  j = \overline{-\infty,\infty},\label{anal10}
\end{align}
with initial condition \eqref{anal_resh4dd}. In this case, relation
(\ref{evolmn-dddd}) becomes
\begin{equation}
\rho(t,s)=\rho_\varphi(s)\exp\bigg[at-\varkappa\sum_{j=-\infty}^{\infty}
\lambda_j v_j(s)\int\limits_{0}^{t}\beta_{j}(t'){\rmd}t'\bigg].\label{anal4y}
\end{equation}

Direct calculation shows that the function $\beta_0(t)$ is a
solution of the logistic equation
\begin{equation}
\dot \beta_0 =a \beta_0-\dfrac{\varkappa}{\sqrt{2\pi}} \lambda_0 \beta_0^2,
\quad  \beta_0\big|_{t=0}=\beta_{00}, \label{anal10q}
\end{equation}
where $\beta_{00}$ is a given constant. So, functions
\begin{equation}
\beta_j(t) = \beta_{0}(t)\delta_{j0} \label{anal_resh4z}
\end{equation}
are solutions of system (\ref{anal10}) with initial conditions
\begin{equation}
\beta_{0j} =\beta_{00}\delta_{j0}.  \label{init_cond}
\end{equation}
From \eqref{anal_resh3b_1} and \eqref{init_cond} we have
\begin{equation}
\rho_\varphi(s)= \sum_{j=-\infty}^{\infty}
\beta_{00}\delta_{j0}v_j(s) = \beta_{00}v_0,
\end{equation}
where $v_0=(\sqrt{2\pi})^{-1}$ is determined by
\eqref{anal_resh3}.
The solution of the Cauchy problems (\ref{anal10q}) has the form
\begin{equation}
\beta_0(t) = \dfrac{\beta_{00}e^{at}}{1+\varkappa\lambda_0\beta_{00}(a\sqrt{2\pi})^{-1}(e^{at}-1)}.  \label{anal_resh4aa}
\end{equation}

Then from \eqref{anal_resh3d} it follows that
\begin{equation}
\rho(t,s)=\rho_0(t,s)=v_0\beta_0(t)=v_0\dfrac{\beta_{00}e^{at}}
{1+\varkappa\lambda_0\beta_{00}(a\sqrt{2\pi})^{-1}(e^{at}-1)}.  \label{anal_resh9s}
\end{equation}
Note that the solution $\rho_0(t,s)$ of the form \eqref{anal_resh9s} is spatially homogeneous.

On the other hand, substituting (\ref{anal_resh4aa}) in \eqref{anal4y}, we also have
\eqref{anal_resh9s}.

\section[The large-time asymptotics]{The large-time asymptotics }

The derivative $(\rho_0)_t(t,s)$ of $\rho_0(t,s)$ from (\ref{anal_resh9s}),
\begin{equation}
(\rho_0)_t(t,s)=\dfrac{v_0\beta_{00}(a-\varkappa\lambda_0v_0\beta_{00})e^{at}}
{[1+\varkappa\lambda_0\beta_{00}(a\sqrt{2\pi})^{-1}(e^{at}-1)]^2},
\label{anal_resh9sa}
\end{equation}
describes the rate of population growth or extinction. According to
\eqref{anal_resh9sa}, the exact solution (\ref{anal_resh9s}) of the
equation (\ref{anal2}) is monotonic in time. The SLD
$\rho_0(t,s)$ increases for $a>\varkappa\lambda_0v_0\beta_{00}$ and
decreases for $a<\varkappa\lambda_0v_0\beta_{00}$. The function
(\ref{anal_resh9s}) tends asymptotically to $\rho_{\lim}$,
\begin{equation}
\rho_{\lim}=\dfrac{a}{\varkappa\lambda_0}
\label{anal_lim}
\end{equation}
as $t\to\infty$. For $a>2\varkappa\lambda_0v_0\beta_{00}$ the
population growth rate reaches a maximum at
\begin{equation}
t_{\max}=\dfrac{1}{a}\ln \bigg(\dfrac{\rho_{\lim}
\sqrt{2\pi}}{\beta_{00}}-1\bigg). \label{anal_resh9sb}
\end{equation}
and then monotonically decreases to zero. For $a<2\varkappa\lambda_0v_0\beta_{00}$
the rate of population growth or extinction is maximal at the initial moment and then
monotonically decreases.

\begin{figure}[h]
\centering
\includegraphics[width=6.4cm,height=4cm]{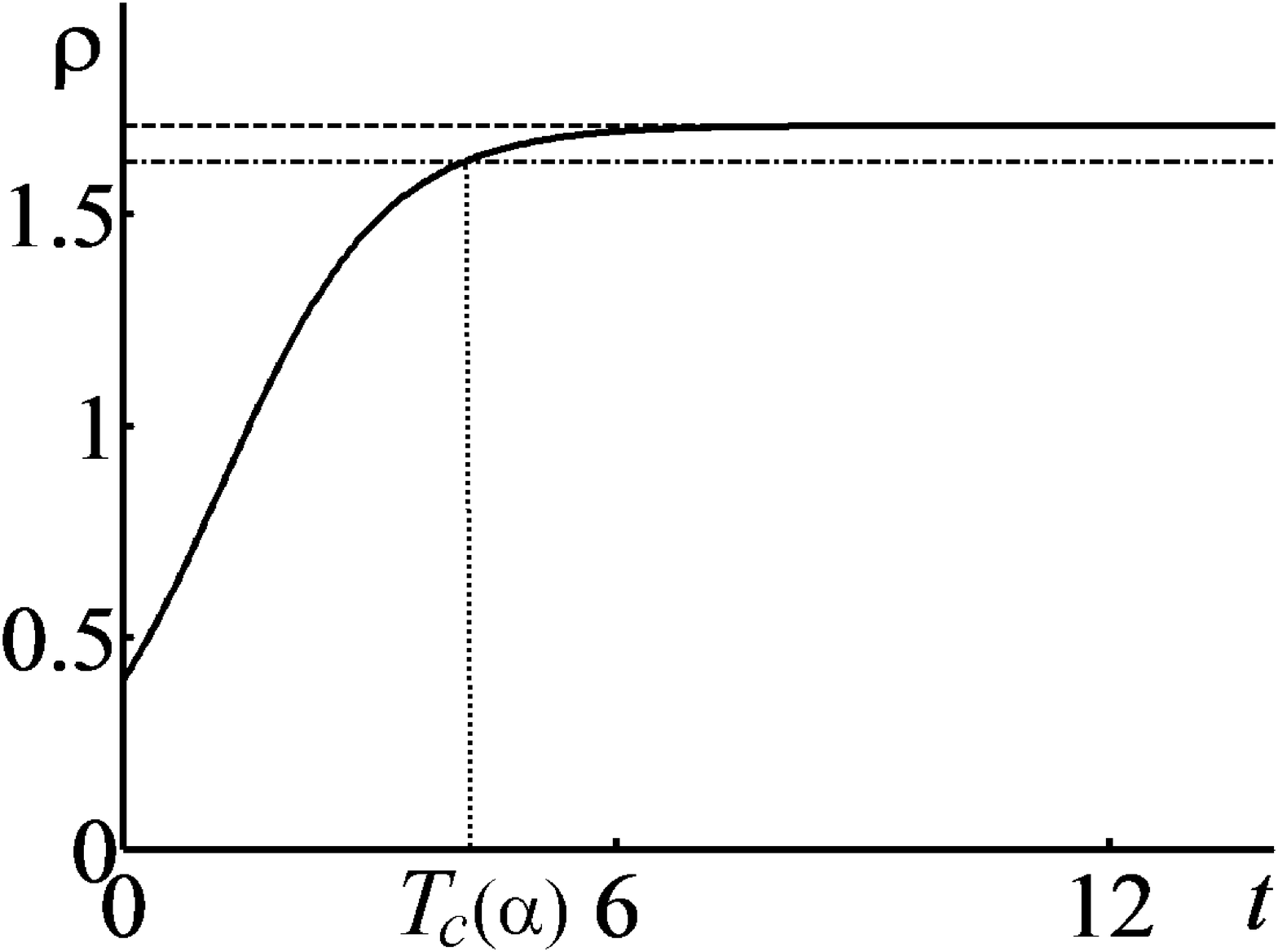} \hfil
\includegraphics[width=6.4cm,height=4cm]{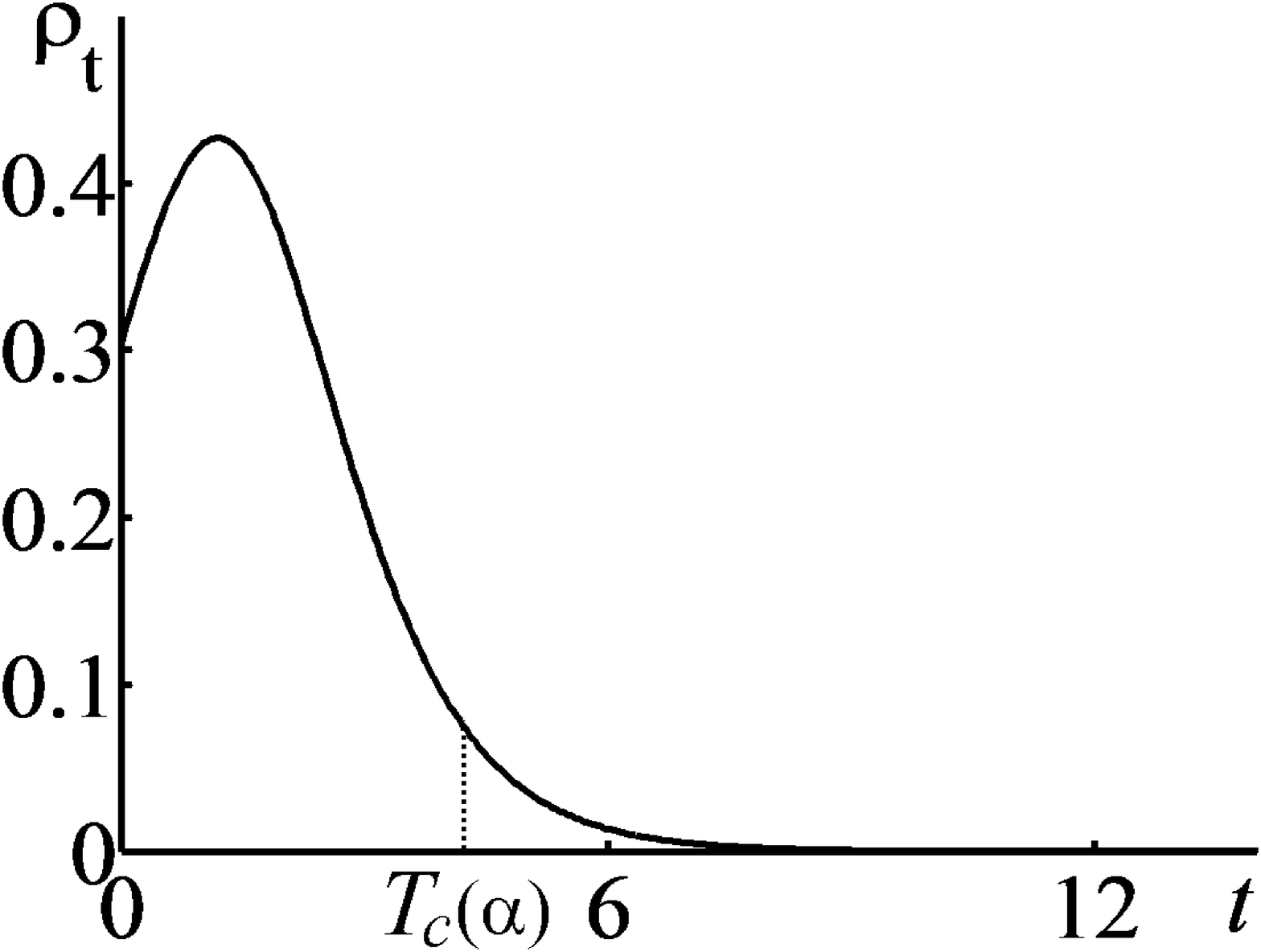} \\
\parbox[t]{.45\textwidth}{\vspace{-5mm}\caption{Graph of the function $\rho_0(t,s)$ for $a=1$,
$a>\varkappa\lambda_0v_0\beta_{00}$,\quad
$\alpha=0.95$}\label{rho1}} \hfil
\parbox[t]{.45\textwidth}{\vspace{-5mm}\caption{Graph of the function $(\rho_0)_t(t,s)$ for  $a=1$,
$a>\varkappa\lambda_0v_0\beta_{00}$,\quad $\alpha=0.95$}\label{rho1p}}\\
\end{figure}

\begin{figure}[h]
\centering
\includegraphics[width=6.4cm,height=4cm]{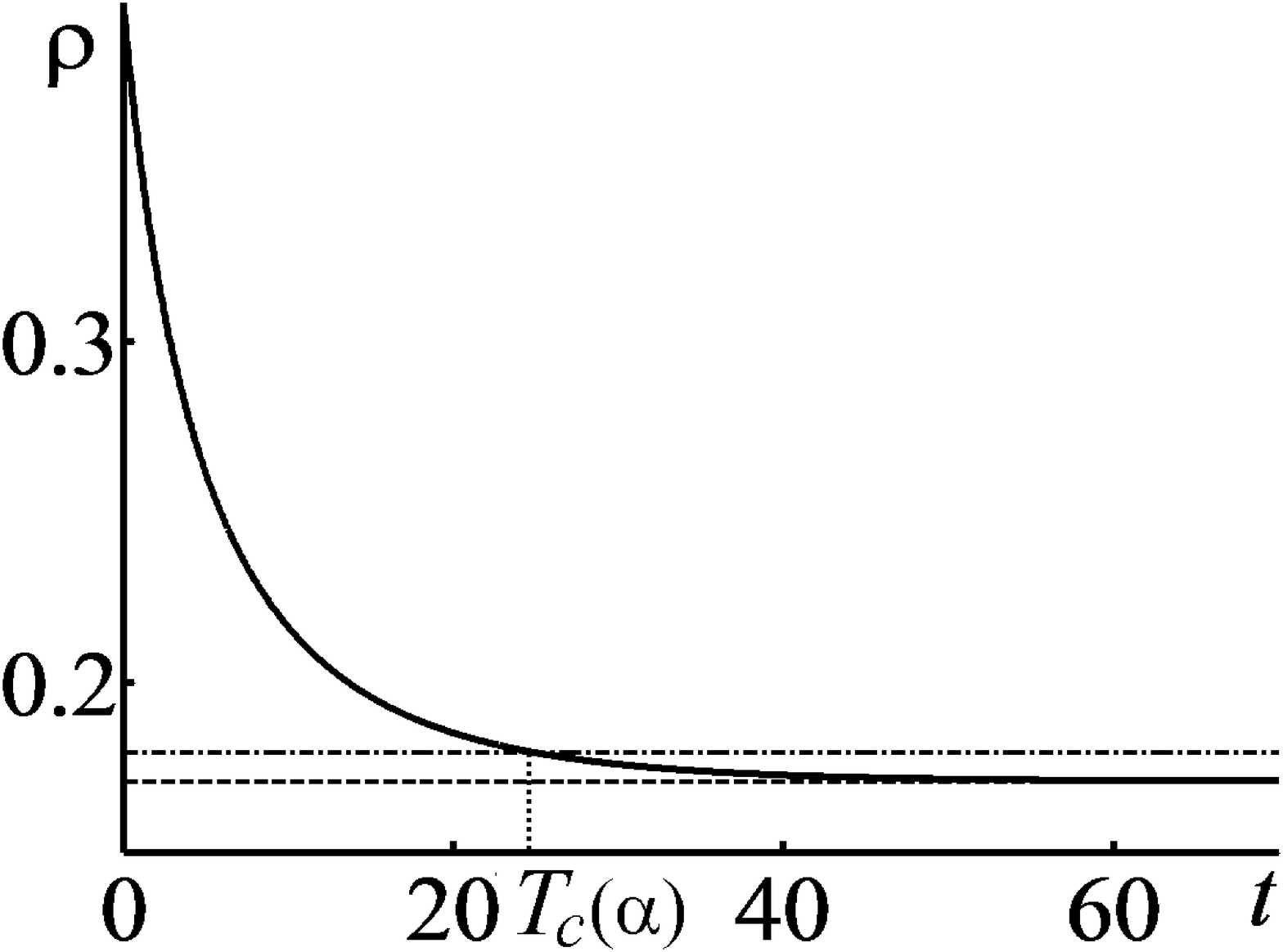} \hfil
\includegraphics[width=6.4cm,height=4cm]{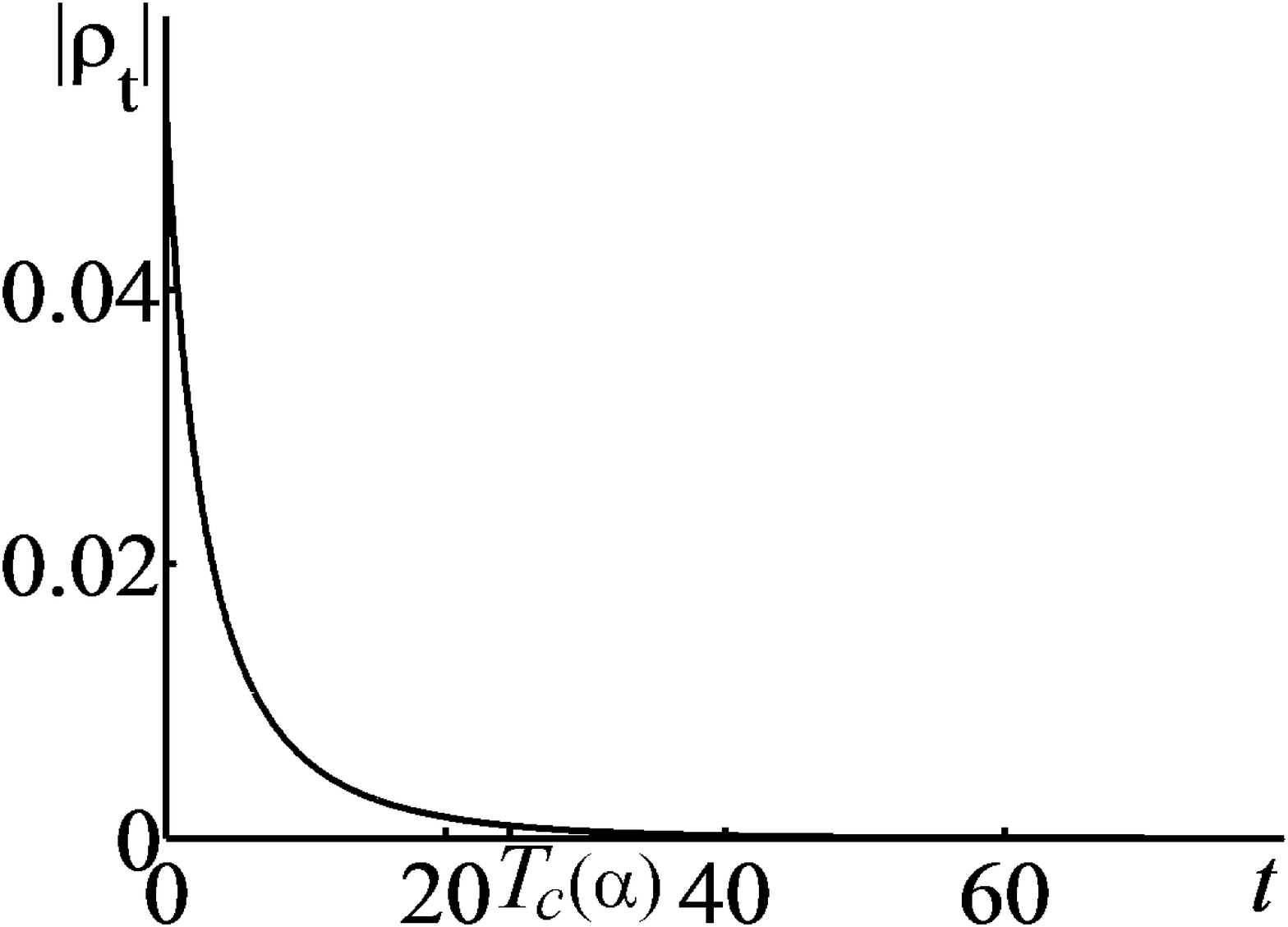} \\
\parbox[t]{.45\textwidth}{\vspace{-5mm}\caption{Graph of the function $\rho_0(t,s)$ for $a=0.1$,
$a<\varkappa\lambda_0v_0\beta_{00}$, \quad
$\alpha=1.05$}\label{rho2}} \hfil
\parbox[t]{.45\textwidth}{\vspace{-5mm}\caption{Graph of the function $|(\rho_0)_t(t,s)|$ for $a=0.1$,
$a<\varkappa\lambda_0v_0\beta_{00}$, \quad $\alpha=1.05$}\label{rho2p}}\\
\end{figure}

Note that $\rho_{\lim}$ given by \eqref{anal_lim} is not a solution
of equation \eqref{anal2}. The solution $\rho_0(t,s)$ from
\eqref{anal_resh9s} can be characterized by time $T_c$, such that
for $t>T_c$ the function $\rho_0(t,s)$ evolves monotonically to the
steady state $\rho_{\lim}$, and the derivative $(\rho_0)_t(t,s)$
\eqref{anal_resh9sa} monotonically tends to zero as $t\to\infty$.
Let us estimate $T_c$ from the condition
\begin{equation}
\Big|\rho_0(t,s)-\rho_{\lim}\Big|<\varepsilon, \quad t>T_c,
\label{dev}
\end{equation}
where $\varepsilon$ is a given small deviation of the solution
$\rho(t,s)$ from the steady-state $\rho_{\lim}$. We consider
\eqref{dev} when the equality is valid and choose the parameter
$\varepsilon$ as
\begin{equation}
\varepsilon=\rho_{\lim}|\alpha-1|.
\end{equation}
The limit $\varepsilon \to 0$ corresponds to $\alpha \to 1$, and
$T_c = T_c(\alpha)$ is defined from the equation
\begin{equation}
\rho_0(T_c(\alpha),s)=\alpha \rho_{\lim}.
\label{anal_resh9sc}
\end{equation}
This equality is illustrated by figures~\ref{rho1} and
~\ref{rho2}. Taking into account the explicit form
\eqref{anal_resh9s} and \eqref{anal_lim} of the functions
$\rho_0(t,s)$ and $\rho_{\lim}$, respectively, we get
\begin{equation}
T_c(\alpha)=\dfrac{1}{a}\ln\bigg\{\dfrac{\alpha}{1-\alpha}
\bigg(\dfrac{\rho_{\lim} \sqrt{2\pi}}{\beta_{00}}-1\bigg)\bigg\}.
\label{anal_resh9sd}
\end{equation}

Figures~\ref{rho1} and ~\ref{rho2} display graphs of the functions
$\rho_0(t,s)$ (solid line), $\rho_{\lim}$ (dashed line) and $\alpha
\rho_{\lim}$ (dash-dot line) for $\alpha = 0.95$, and
$\alpha = 1.05$, respectively, for $b_0=1$, $\varkappa=0.2$, $R=1$,
$\gamma=1$, $\rho_\varphi = 1/\sqrt{2\pi}$. Figures~\ref{rho1p} and
~\ref{rho2p} display graphs of the derivative modules
$|(\rho_0)_t(t,s)|$ for $\alpha = 0.95$, and $\alpha = 1.05$,
respectively, and for the same values of the equation parameters. We
see that for $t> T_c(\alpha)$ the function $\rho_0(t,s)$ monotonically
tends to $\rho_{\lim}$, and the function $|(\rho_0)_t(t,s)|$
monotonically tends to zero. According to \cite{cunha2009,fuentes2003},
the parameters of equation \eqref{anal2} are considered to be
non-dimensional.

Solutions of equation \eqref{anal2} tend to a steady-state at large times
\cite{naumkin1994,shismarev1999,komarov2001,komarov2002,komarov2011}.
This property comes to the idea to seek solutions of equation (\ref{anal2}) in a class of
functions closely similar to (\ref{anal_resh9s}).


Denote by $T$ ($T > T_c(\alpha)$) a large parameter that
characterizes the evolution time for equation \eqref{anal2}. Let
$t=T\tau$ where $\tau \in [0,1]$. Denote ${\mathcal K}_T^t={\mathcal K}_T^t(\phi)$
class of functions of the form
\begin{equation}
{\mathcal K}_T^t = \bigg\{\beta(t) \bigg| \beta(t) = \beta(\theta,\tau,T)
= \beta^{(0)} (\theta,\tau)+
\dfrac{1}{T}\beta^{(1)} (\theta,\tau)+\ldots, \quad \theta = T
\phi(\tau) \bigg\}. \label{vst_beta1}
\end{equation}
Here, the function $\phi(\tau)$ is a functional parameter of the
class ${\mathcal K}_T^t$. We refer to the variables $\theta$ and $\tau$ as a
fast and a slow variables, respectively.

Let us seek solutions of the system \eqref{anal10} in the class of
functions ${\mathcal K}_T^t$. In view of \eqref{anal_resh3d} and
\eqref{vst_beta1} we obtain
\begin{equation}
\rho(t,s) = \rho(\theta,\tau,s) = \rho^{(0)} (\theta,\tau,s) +
\dfrac{1}{T} \rho^{(1)} (\theta,\tau,s)+\ldots, \quad T \to \infty.
\label{anal_resh15}
\end{equation}

Choose the initial condition for the function $\rho(t,s)$ as
\begin{equation}
\rho_\varphi(s)=\beta_{00}v_0 + \dfrac1T\tilde\rho_\varphi(s).
\label{vst_beta0}
\end{equation}

We refer to solution of the form \eqref{anal_resh15} of equation
\eqref{anal2} as \emph{quasi-steady-state solution}. It is important that
the function $\rho^{(0)} (\theta,\tau,s)$ determines behavior of the
quasi-steady-state solution as $T \to \infty$.

The slow variable $\tau$ and the fast variable $\theta$ scale such that
\begin{equation}
\dfrac{d}{dt} = \dfrac{\partial\theta}{\partial t} \dfrac{\partial}{\partial\theta}+ \dfrac{{\rmd}\tau}{{\rmd}t} \dfrac{\partial}{\partial\tau} = \phi_\tau
\dfrac{\partial}{\partial\theta} + \dfrac{1}{T}\dfrac{\partial
}{\partial\tau}.\label{anal_resh16}
\end{equation}
Then (\ref{anal10}) reads
\begin{align}
& \bigg[\phi_\tau \dfrac{\partial}{\partial\theta} + \dfrac{1}{T}\dfrac{\partial}{\partial\tau}\bigg]
  \bigg(\beta^{(0)}_j + \dfrac{1}{T}\beta^{(1)}_j+\ldots\bigg) = a\bigg(\beta^{(0)}_j + \dfrac{1}{T}\beta^{(1)}_j+\ldots\bigg)- \nonumber\\
&\qquad - \dfrac{\varkappa}{\sqrt{2\pi}} \sum_{l=-\infty}^{\infty}
\lambda_l \bigg(\beta^{(0)}_{j-l} +
\dfrac{1}{T}\beta^{(1)}_{j-l}+\ldots\bigg)\bigg(\beta^{(0)}_{l} +
\dfrac{1}{T}\beta^{(1)}_{l}+\ldots\bigg).\label{vst_beta2}
\end{align}

Equating terms involving the same power $1/T$, we obtain
\begin{align}
\dfrac{1}{T^0}:\; & \phi_\tau \dfrac{\partial}{\partial\theta} \beta^{(0)}_j
=a\beta^{(0)}_j -\dfrac{\varkappa}{\sqrt{2\pi}}
\sum_{l=-\infty}^{\infty} \lambda_l\beta^{(0)}_{j-l}\beta^{(0)}_{l},\label{vst_beta3b}\\
\dfrac{1}{T^1}:\; & \phi_\tau \dfrac{\partial}{\partial\theta} \beta^{(1)}_j= a\beta^{(1)}_j -
\dfrac{\varkappa }{\sqrt{2\pi}} \sum_{l=-\infty}^{\infty}\lambda_l
\big(\beta^{(1)}_{j-l} \beta^{(0)}_{l} + \beta^{(0)}_{j-l} \beta^{(1)}_{l}\big)
-\dfrac{\partial}{\partial\tau} \beta^{(0)}_j , \dots\label{vst_beta3c}
\end{align}

Equation (\ref{vst_beta3b}) for $j=0$ reads
\begin{equation}
\phi_\tau^{(0)} \dfrac{\partial}{\partial\theta} \beta^{(0)}_0 =
a\beta^{(0)}_0 -\dfrac{\varkappa }{\sqrt{2\pi}}
\sum_{l=-\infty}^{\infty} \lambda_l \beta^{(0)}_{-l}\beta^{(0)}_{l}.
\label{vst_beta3a}
\end{equation}

With the initial conditions \eqref{vst_beta0}, we obtain the following solution of equations (\ref{vst_beta3b}) and (\ref{vst_beta3a}):
\begin{eqnarray}
& \beta_0^{(0)}(\theta,\tau)=
\dfrac{\beta_{00}{\rme}^{a\theta/\phi_\tau}}
{1+\varkappa\lambda_0\beta_{00}(a\sqrt{2\pi})^{-1} ({\rme}^{a\theta/\phi_\tau}-1)},\label{anal_resh19aa}\\
& \beta_j^{(0)}(\theta,\tau) = 0, \quad j \neq 0.\nonumber
\end{eqnarray}

Without loss of generality, we have (see, e.g.,
\cite{Maslov2,litvinets2006})
\begin{equation}
\phi(\tau) = a\tau.\label{anal_resh20}
\end{equation}
Then (\ref{anal_resh19aa}) takes the form
$\beta^{(0)}_0(t) = \beta_0(t)$ where $\beta_0(t)$ is given by
\eqref{anal_resh4aa}. Similarly, from \eqref{anal_resh3d}, we
obtain: $\rho^{(0)}(t,s)=\rho_0(t,s)$. Here, $\rho_0(t,s)$ is determined by
(\ref{anal_resh9s}). Then an asymptotic solution $\rho(t,s)$ of equation
\eqref{anal2} is
\begin{equation}
\rho(t,s) = \rho^{(0)}(t,s)\big[1+O(1/T)\big],  \quad T\to \infty.
\label{anal_resh9s_ind01}
\end{equation}

In analogy with \cite{cunha2009,cunha2011}, where a
one-dimensional pattern is considered as a perturbation of the exact
steady-state solution of the FKPP equation \eqref{intr1}, we
describe patterns as perturbations of exact non-steady-state
solution \eqref{anal_resh9s} of equation \eqref{anal2}.

We refer to patterns of such type \emph{as quasi-steady-state patterns}.
They evolve monotonically to the steady-state $\rho_{\lim}$ given by
\eqref{anal_lim} (as $T \to \infty$).

From (\ref{anal_resh19aa}) it follows that
$$
\dfrac{\partial}{\partial\tau} \beta^{(0)}_j = 0.
$$

We write equation (\ref{vst_beta3c}) as
\begin{eqnarray}
&& \dfrac{\partial}{\partial\theta} \beta^{(1)}_0 = \beta^{(1)}_0 - \dfrac{2\varkappa \lambda_0}{a\sqrt{2\pi}} \beta^{(0)}_{0} \beta^{(1)}_{0},\nonumber\\
&& \dfrac{\partial}{\partial\theta} \beta^{(1)}_j = \beta^{(1)}_j
-\dfrac{\varkappa }{a \sqrt{2\pi}} \beta^{(0)}_{0} \big(\lambda_0
\beta^{(1)}_{j} + \lambda_{j} \beta^{(1)}_{j}\big),\label{vst_beta4}\\
&& \dfrac{\partial}{\partial\theta} \beta^{(1)}_{-j} = \beta^{(1)}_{-j}
-\dfrac{\varkappa}{a \sqrt{2\pi}} \beta^{(0)}_0 \big(\lambda_0
\beta^{(1)}_{-j} + \lambda_{-j} \beta^{(1)}_{-j}\big).\nonumber
\end{eqnarray}

If the initial distribution \eqref{vst_beta0} is symmetric, $\lambda_{-j}=\lambda_{j}$, then the equations in $\beta^{(1)}_{j}$ is identical to that in $\beta^{(1)}_{-j}$ and reads
\begin{equation}
\dfrac{\partial}{\partial\theta} \beta^{(1)}_j = \beta^{(1)}_j -
\dfrac{\varkappa(\lambda_j+\lambda_0)}{a\sqrt{2\pi}} \beta^{(0)}_0
\beta^{(1)}_{j}. \label{vst_beta5}
\end{equation}

Solution of system (\ref{vst_beta4}) is
\begin{align}
\beta^{(1)}_0(\theta,\tau)&=\dfrac{\beta_{10}{\rme}^{\theta}}
{[1+\varkappa\lambda_0\beta_{00}(a\sqrt{2\pi})^{-1}({\rme}^{\theta}-1)]^2}
= \dfrac{\beta_{10}{\rme}^{at}}{[1+\varkappa\lambda_0\beta_{00}(a\sqrt{2\pi})^{-1}(\rme^{at}-1)]^2}, \nonumber\\
\beta^{(1)}_j(\theta,\tau)&=
\dfrac{\beta_{1j}{\rme}^{\theta}}{[1+\varkappa\lambda_0\beta_{00}
(a\sqrt{2\pi})^{-1}({\rme}^{\theta}-1)]^{(\lambda_j+\lambda_0)/\lambda_0}}=
\nonumber\\
&=\dfrac{\beta_{1j}{\rme}^{at}}{[1+\varkappa\lambda_0\beta_{00}
(a\sqrt{2\pi})^{-1}({\rme}^{at}-1)]^{(\lambda_j+\lambda_0)/\lambda_0}},
\label{vst_beta6}
\end{align}
where
\begin{equation}
\beta_{1j}=\dfrac1{\sqrt{2\pi}}\int\limits_{-\pi}^\pi
   \tilde\rho_\varphi(s){\rme}^{-ijs}{\rmd}s. \label{ann}
\end{equation}
Here, $\tilde \rho_\varphi(s)$ is given by \eqref{vst_beta0}.

Then, for the case of a symmetric initial density distribution, we
get
\begin{align}
\rho(t,s)&= \rho^{(0)}(t,s)+\dfrac{1}{T} \rho^{(1)}(t,s)=
  v_0 \dfrac{\beta_{00} {\rme}^{at}}{1+\varkappa\lambda_0\beta_{00}(a\sqrt{2\pi})^{-1}
  ({\rme}^{at}-1)} + \nonumber\\
&+\dfrac{1}{T\sqrt{2\pi}}
\sum_{j=-\infty}^{\infty}\dfrac{\beta_{1j}{\rme}^{at}
{\rme}^{ijs}}{[1+\varkappa\lambda_0\beta_{00}(a\sqrt{2\pi})^{-1}
({\rme}^{at}-1)]^{1+I_j(\mu)/I_0(\mu)}}+\ldots.\label{vst_beta8}
\end{align}

Note, that substitution of \eqref{vst_beta1} directly in
(\ref{anal4y}) and in view of \eqref{anal_resh19aa} and
\eqref{vst_beta6} also yields (\ref{vst_beta8}) (see Appendix A).

The proposed procedure for constructing asymptotic solutions can be
applied immediately to equation (\ref{anal2}). Representing the
function $\rho(t,s)$ as \eqref{anal_resh15},
substituting (\ref{anal_resh15}) in (\ref{anal2}) and taking into
account (\ref{anal_resh16}), we have
\begin{multline}
\bigg[\phi_\tau^{(0)}\dfrac{\partial}{\partial\theta} + \dfrac{1}{T}\dfrac{\partial}{\partial\tau}\bigg]
  \bigg(\rho^{(0)}+\dfrac{1}{T}\rho^{(1)}+\ldots\bigg)=\\
=a\bigg(\rho^{(0)} + \dfrac{1}{T}\rho^{(1)}+\ldots\bigg)-\varkappa
  \bigg(\rho^{(0)} + \dfrac{1}{T}\rho^{(1)}+\dots\bigg) \times\\
\times \int\limits_{-\pi}^{\pi}
\tilde b_\gamma(s,s')\bigg(\rho^{(0)}(\theta,\tau,s')+
  \dfrac{1}{T}\rho^{(1)}(\theta,\tau,s')+\ldots\bigg){\rmd}s'.
\label{anal_resh17}
\end{multline}

Equating terms involving the same powers of $1/T$, we obtain
\begin{align}
\phi_\tau^{(0)} \rho^{(0)}_\theta&= a\rho^{(0)}-\varkappa\rho^{(0)}\int\limits_{-\pi}^{\pi} \tilde b_\gamma(s,s')\rho^{(0)}(s'){\rmd}s', \label{anal_resh18a} \\
\phi_\tau^{(0)} \rho^{(1)}_\theta&= a\rho^{(1)} - \varkappa\rho^{(1)}\int\limits_{-\pi}^{\pi} \tilde b_\gamma(s,s')\rho^{(0)}(s'){\rmd}s' - \nonumber\\
& -\varkappa\rho^{(0)}\int\limits_{-\pi}^{\pi} \tilde b_\gamma (s,s')\rho^{(1)}(s'){\rmd}s'- \rho^{(0)}_\tau,\label{anal_resh18b} \\
\dots& \dots \nonumber
\end{align}

According to Section 4, for initial condition \eqref{vst_beta0} we obtain
$\rho^{(0)}(\theta,\tau,s) = \rho^{(0)}(t,s)=\rho_0(t,s)$ where $\rho_0(t,s)$
is given by (\ref{anal_resh9s}). If we seek a solution to equation
\eqref{anal_resh18b} as an expansion in the eigenfunctions of the kernel \eqref{anal_resh3}
\begin{equation}
\rho^{(1)} (\theta,\tau,s)=
\sum_{j=-\infty}^{\infty}C_j(\tau,\theta)v_j(s),\label{anal_resh21}
\end{equation}
we also get (\ref{vst_beta8}) (see Appendix B).

\begin{figure}[h]
\centering
\includegraphics[width=6.4cm,height=3.8cm]{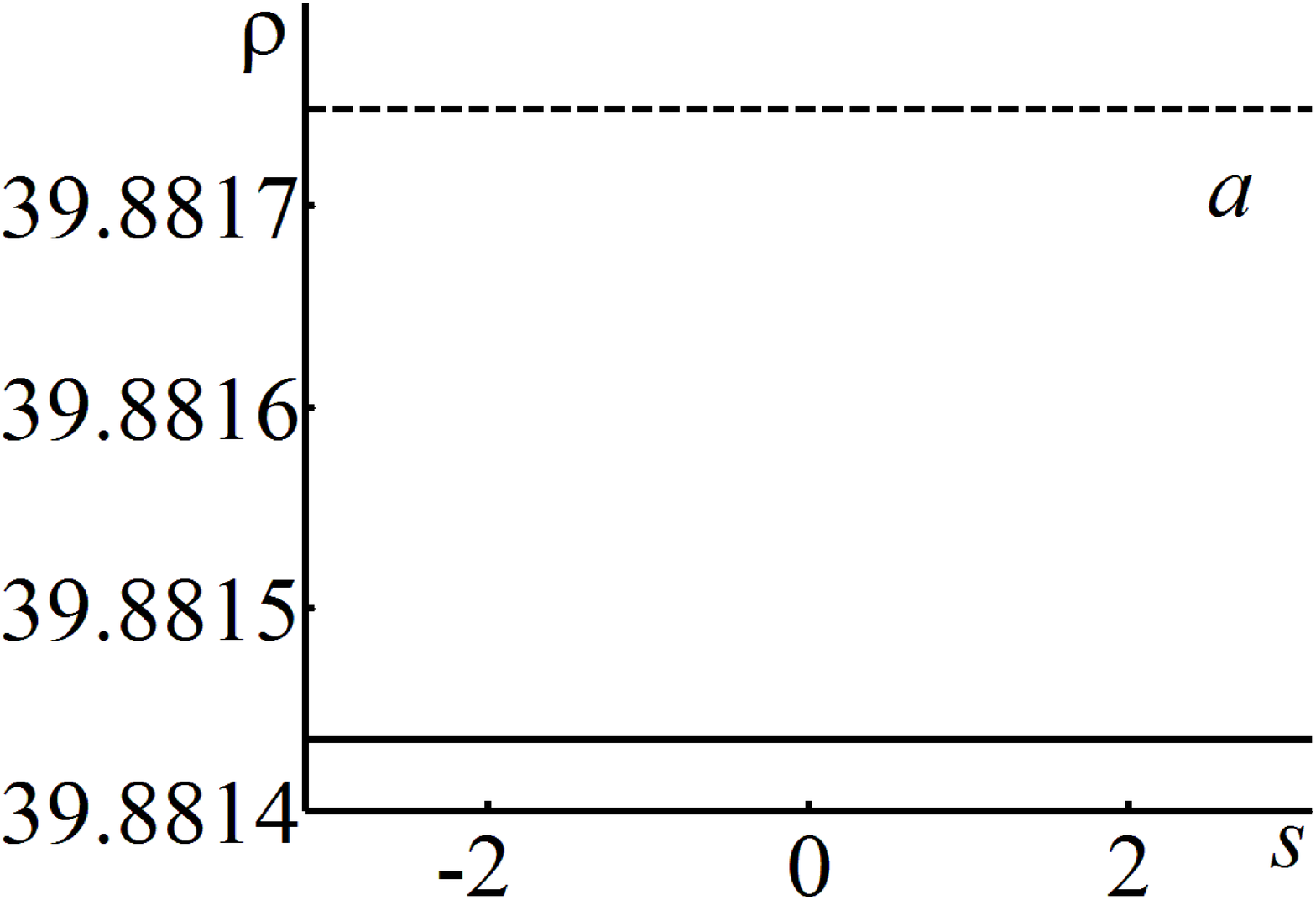} \hfil
\includegraphics[width=6.4cm,height=3.8cm]{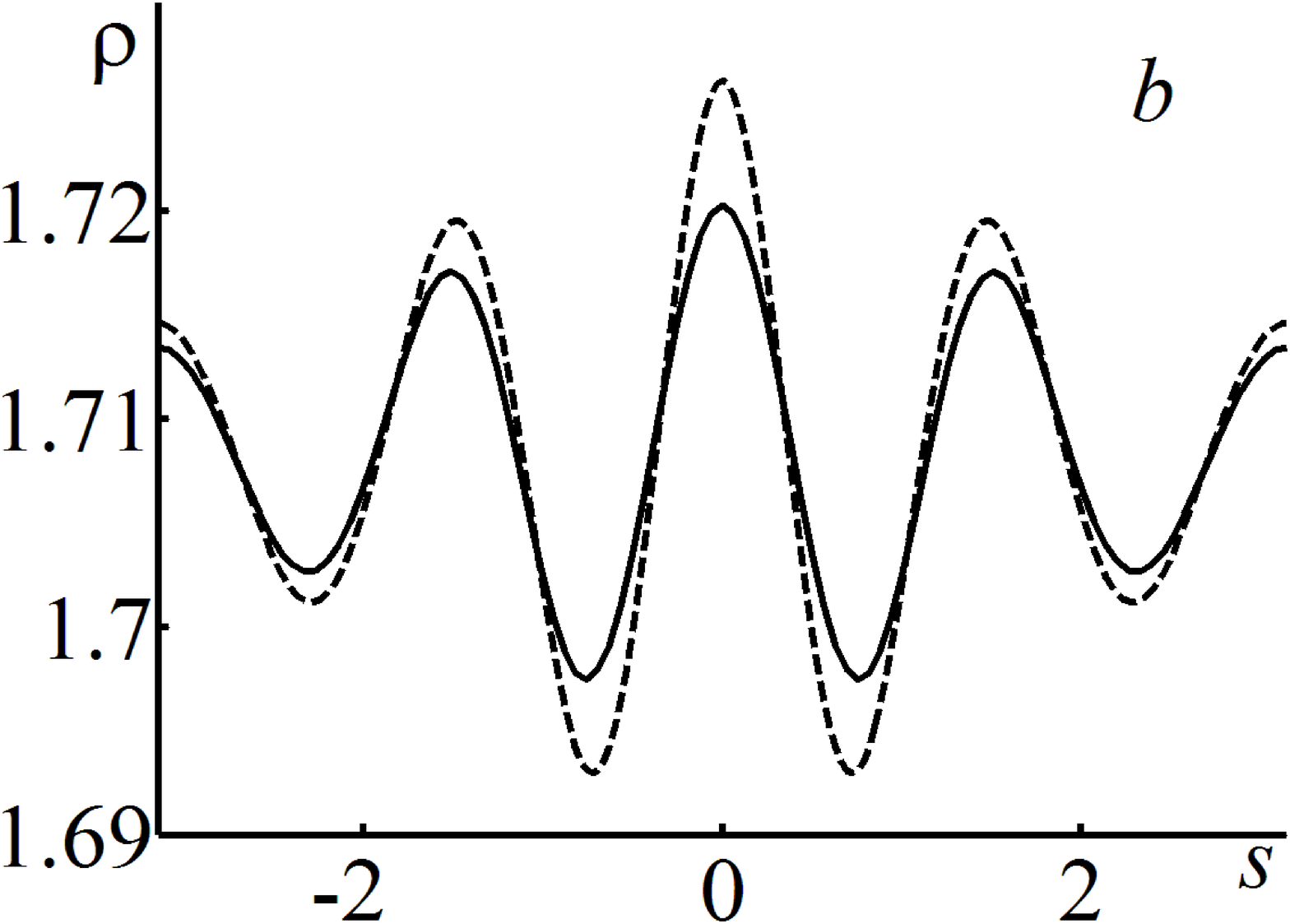} \\
\includegraphics[width=6.4cm,height=3.8cm]{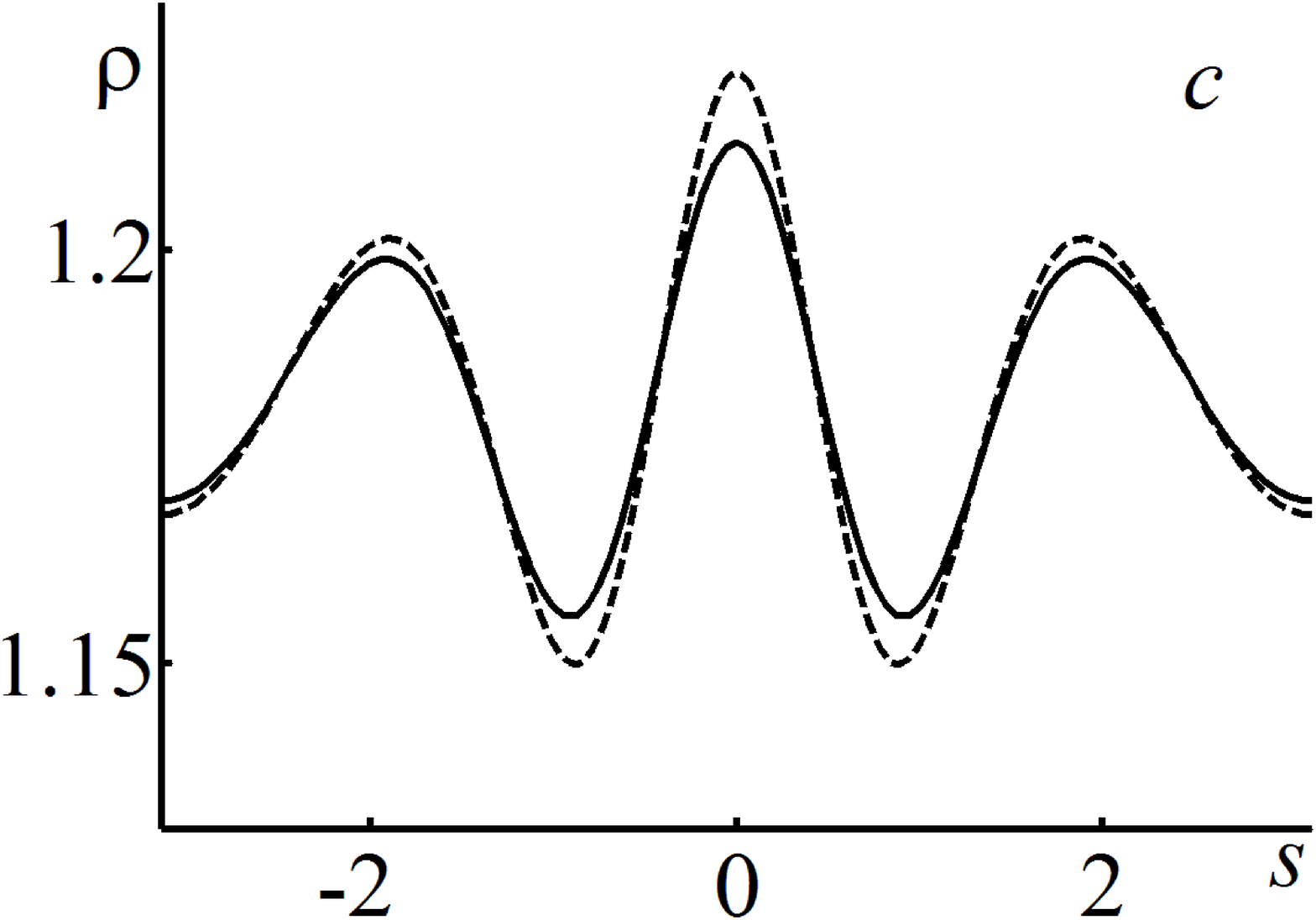}\hfil
\includegraphics[width=6.4cm,height=3.8cm]{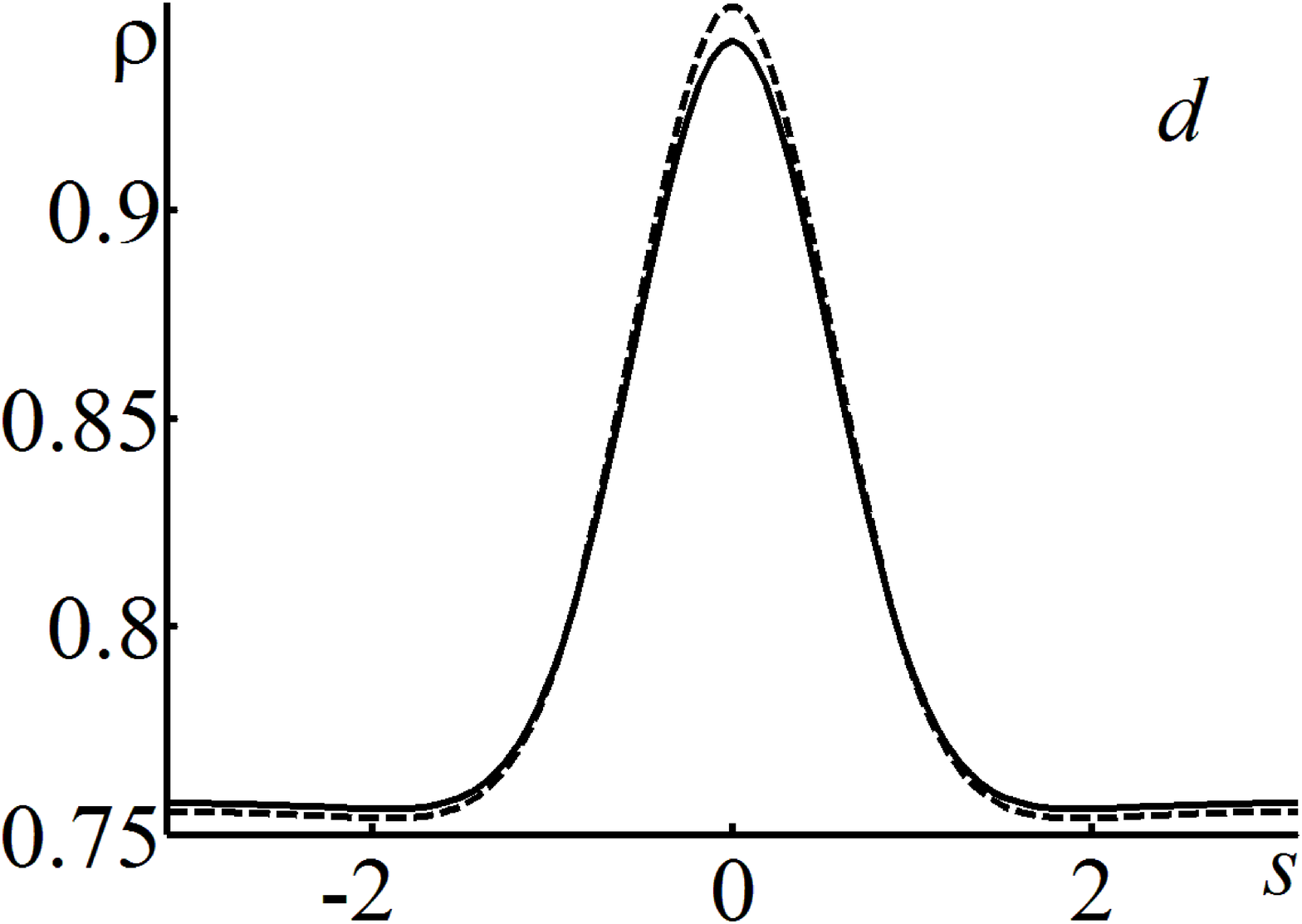}
\caption{Graph of the function $\rho(t,s)$ for $t=200$
and $\gamma = 0.05$ ($a$), 1 ($b$), 1.5 ($c$), and 50 ($d$). The
solid curve shows $\rho(t,s)$ obtained by numerical simulation, the
dashed curve shows the analytical solution $\rho(t,s)$ given by
\eqref{vst_beta8} \label{an_chis}}
\end{figure}

To verify asymptotic formula \eqref{vst_beta8},
we have carried out direct numerical simulation of equation \eqref{anal2}.
In (\ref{vst_beta8}) we have set $j=10$, i.e. we have taken into account 21 terms.
We have used known difference scheme for updating function $\rho(t,s)$
\cite{hoffman}. Figure \ref{an_chis} displays the SLD $\rho(t,s)$
obtained by equation (\ref{vst_beta8}) (dashed line) and by numerical
simulations (solid line) for $a=1$, $b_0=1$, $\varkappa=0.2$,
$D=0.1$, $V=0$, $W=0$, $R=1$, $T=10$, $\rho_\varphi(s) =
(\sqrt{2\pi})^{-1}+T^{-1}\exp(-s^2/0.6)$.

When $\gamma \ll R$, the function $\tilde b_\gamma(s,s')$ of the
form (\ref{anal4a}) is a $\delta$-shaped sequence which tends to the
$\delta$-function as $\gamma \to 0$. In this case, equation \eqref{anal2}
becomes local and patterns do not form (Fig.~\ref{an_chis}$a$).
If $\gamma \gg R$, the function $\tilde b_\gamma(s,s')$ can be
represented asymptotically: $\tilde b_\gamma(s,s')\approx b_0$ as
$\gamma\to\infty$. Then equation \eqref{anal2} tends to the form
\begin{equation}
\rho_t(t,s)= a\rho(t,s)- \sqrt{2\pi} \varkappa b_0 \rho(t,s) \beta_0(t)
\end{equation}
where $\beta_0(t)$ is defined in \eqref{anal_resh4d}. Patterns are also not
observed (Fig.~\ref{an_chis}$d$). Figures~\ref{an_chis}$b$ and~\ref{an_chis}$c$
display patterns for $t=200$. Note that the number of peaks formed decreases with
increasing $\gamma$ for a given $R$.

In the early evolution formation of additional peaks occurs.
Then the formed structure evolves steadily without changing its quality,
i.e. becomes quasi-stady-state.

\begin{figure}[h]
\centering
\includegraphics[width=6.02cm,height=4.21cm]{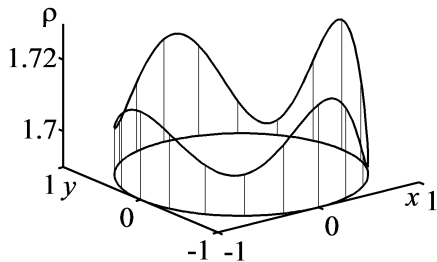}
\caption{Graph of the function $\rho(t,s)$ on the manifold $\Lambda_t^1$
} \label{ob}
\end{figure}

As can be seen from Fig.~\ref{an_chis}, the numerical and analytical
approaches are in good agreement. Note that all graphs in Fig.~\ref{an_chis}
are one-dimensional sweeps of the 2D truncated distributions. For
clarity, we represent pattern displayed on Fig.~\ref{an_chis}$b$
(dashed line) in 3D space (see Fig.~\ref{ob}).

\section{The equation for SLD with diffusion}

\begin{figure}[h]
\centering
\includegraphics[width=6.4cm,height=3.4cm]{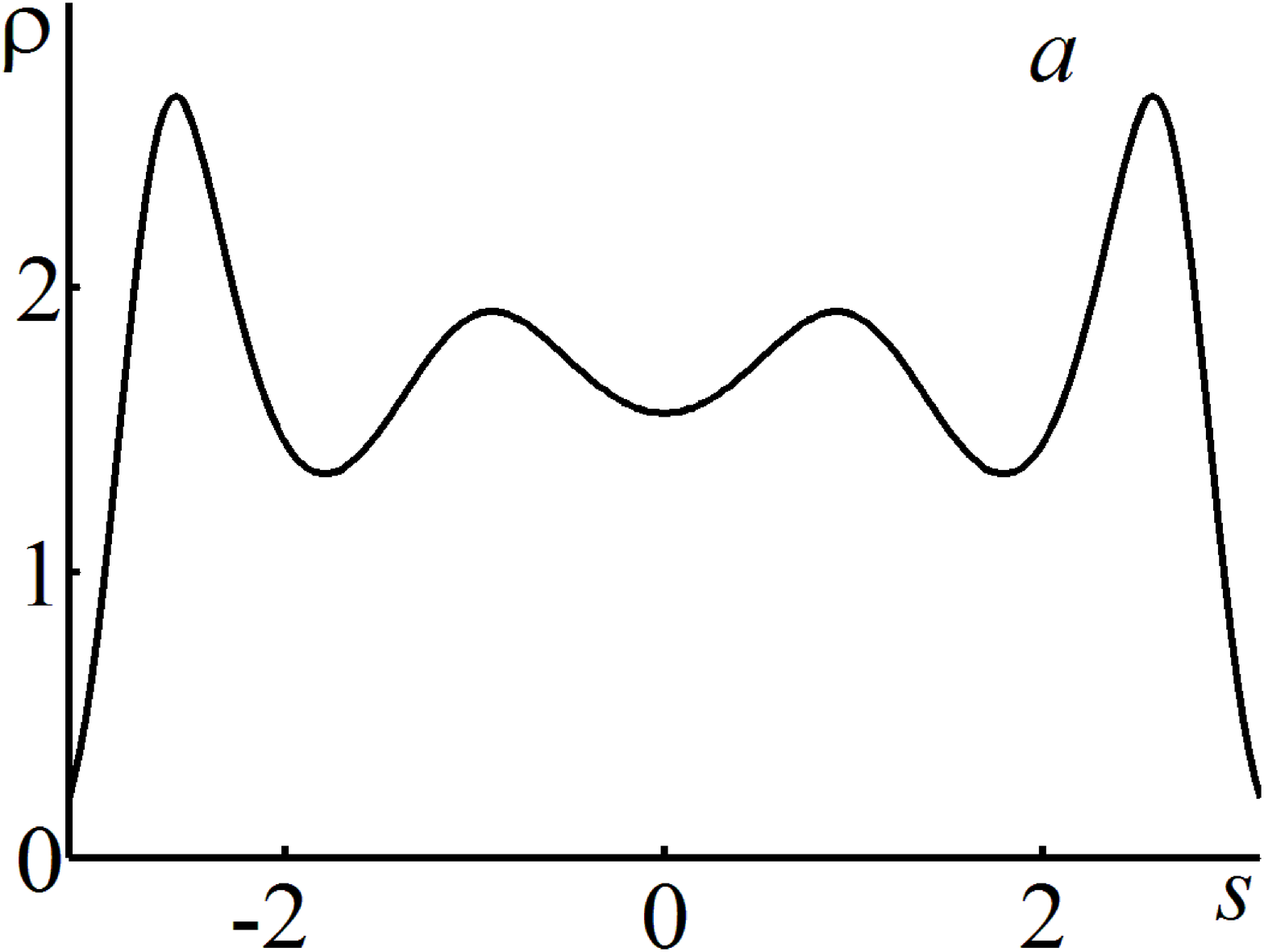}\hfil
\includegraphics[width=6.4cm,height=3.4cm]{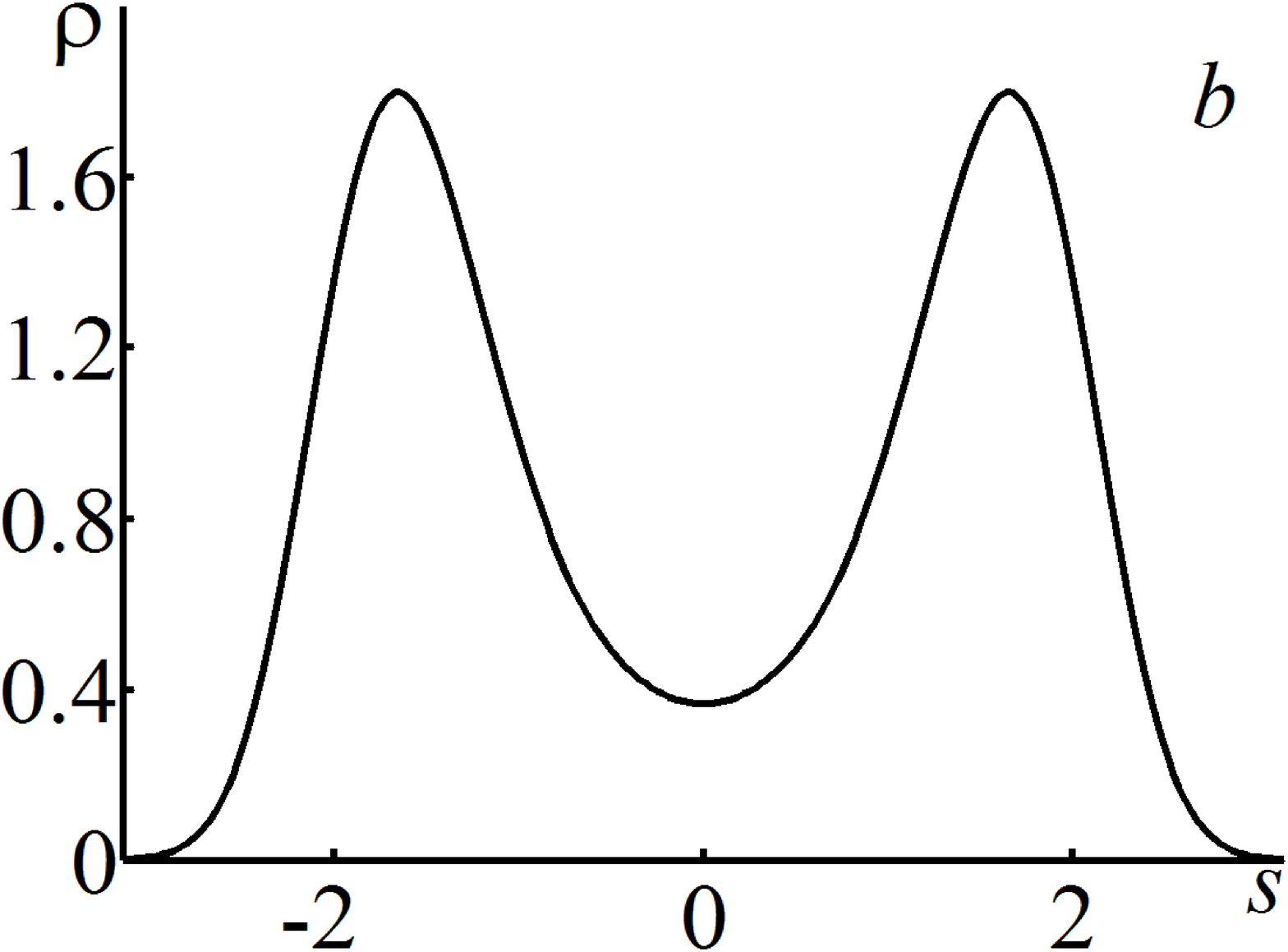}
\caption{Graphs of the function $\rho(t,s)$ obtained by numerical
simulation for $\gamma = 1$, $a=1$,  $b_0=1$, $\varkappa=0.2$,
$D=0.1$, $W=0$, $R=1$, $\rho_\varphi(s) = \exp(-s^2/0.6)$ and
$k_0=0$ ($a$), $k_0=0.03$ ($b$) \label{s_bez}}
\end{figure}

Equation (\ref{evolmn-dd}) does not account the diffusion and the
transport effects while the models with either diffusion or convection
considered in \cite{fuentes2003,cunha2009} predicted numerically
occurrence of patterns. Note, that the convection term
($V_{\vec x}(\vec x,t) \neq 0$) involved in equation \eqref{eq1} affects lower
dimensional patterns, which are qualitatively different from the
full-sized patterns considered in e.g., \cite{cunha2009,cunha2011}.
Addition of the linear convection term, $V_{\vec x}=-k_0\vec x$, to equation \eqref{eq1} results
in a compression of the original manifold, according to (\ref{evolmn2}).
Therefore, density peaks are formed in smaller number at the same
values of $\gamma$ (Fig.~\ref{s_bez}). It has been found
\cite{cunha2009}, that convection is responsible for the motion of
the population in space. For lower dimensional patterns,
convection affects mainly the evolution of the original manifold.

Let us next investigate how diffusion affects the pattern formation
by adding a diffusion term to equation (\ref{evolmn-dd}). Thus we have
the equation
\begin{gather}
\rho_t(t,s)=D \rho_{ss}(t,s)+a\rho(t,s)- \varkappa \rho(t,s)
\int\limits_{-\pi}^\pi \tilde b_\gamma(s,s')\rho(s',t){\rmd}s', \nonumber\\
D={\rm const}, \quad a(\vec x,t)=a={\rm const}, \label{1d_1}
\end{gather}
with the boundary and the initial conditions, respectively
$$
\rho(0,s)=\rho_\varphi(s), \quad \rho(t,s+2\pi) = \rho(t,s).
$$

The eigenfunctions and eigenvalues of the Fredholm operator with the
kernel $\tilde b(s,s')$ are defined by \eqref{anal4a}. We will
seek a solution to equation \eqref{1d_1} in the form \eqref{anal_resh3d}
where $v_j(s)$ are the eigenfunctions of the kernel $\tilde
b(s,s')$, and the Fourier coefficients $\beta_j(t)$
are calculated by (\ref{anal_resh4d}). In view of
\eqref{anal_resh3d}, equation \eqref{1d_1} can
be written as
\begin{equation}
\rho_t(t,s)=D \rho_{ss}(t,s)+a\rho(t,s)- \varkappa
\rho(t,s)\bigg\{\sum_{p=-\infty}^{\infty}\lambda_p\beta_{p}(t)v_p(s)\bigg\}
.\label{anal4x_1}
\end{equation}

As a result, system (\ref{anal_resh4dd1}) becomes
\begin{equation}
\dot \beta_j =\tilde a_j \beta_j-\dfrac{\varkappa}{\sqrt{2\pi}} \sum_{p=-\infty}^{\infty}\lambda_p \beta_{j-p}\beta_{p},\quad  j = \overline{-\infty,\infty},\label{anal10_1}
\end{equation}
where
\begin{equation}
\tilde a_j = -Dj^2+a \label{a_j}
\end{equation}
and the initial condition $\beta_{0j}$ is determined by \eqref{anal_resh4dd}.

Repeating the reasoning used in the previous sections, we obtain for
the initial distribution \eqref{vst_beta0}
\begin{align}
\rho(t,s)&=v_0 \dfrac{\beta_{00}
{\rme}^{at}}{1+\varkappa\lambda_0\beta_{00}(a\sqrt{2\pi})^{-1}({\rme}^{at}-1)}
+ \nonumber\\
& +\dfrac{1}{T\sqrt{2\pi}}
\sum_{j=-\infty}^{\infty}\dfrac{\beta_{1j}{\rme}^{\tilde a_j t}
{\rme}^{ijs}}{[1+\varkappa\lambda_0\beta_{00}(a\sqrt{2\pi})^{-1}
({\rme}^{at}-1)]^{1+I_j(\mu)/I_0(\mu)}}+\ldots\label{vst_beta8a}
\end{align}
where $\beta_{1j}$ is given by \eqref{ann}.

From \eqref{vst_beta8a} we see that as $D$ increases, then
$\tilde a_j$ given by (\ref{a_j}) decreases, and patterns do not form.
This fact is in good agreement  with the results reported in \cite{kenkre2004}.

\begin{figure}[h]
\centering
\includegraphics[width=6.4cm,height=3.4cm]{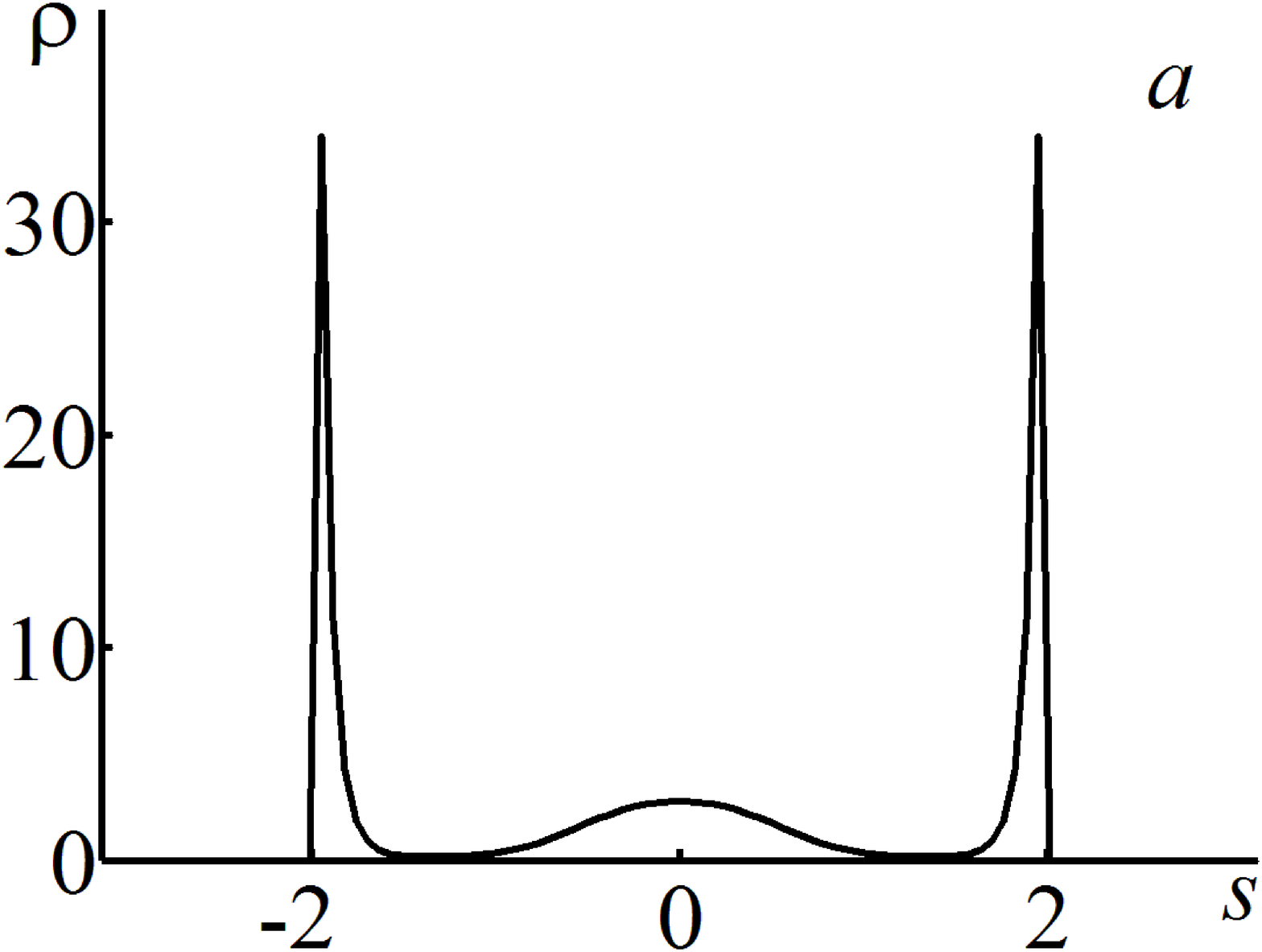}\hfil
\includegraphics[width=6.4cm,height=3.4cm]{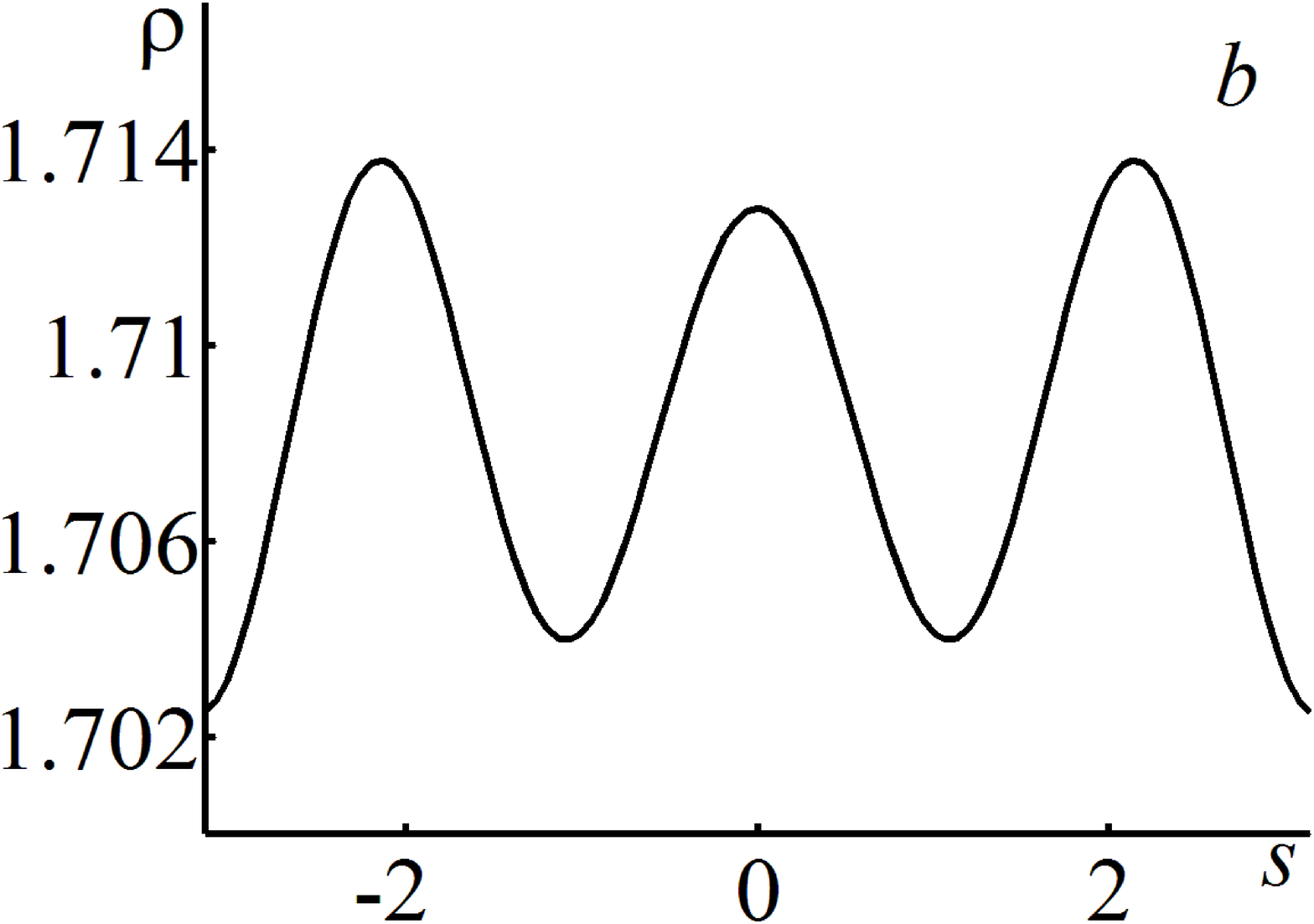} \\
\includegraphics[width=6.4cm,height=3.4cm]{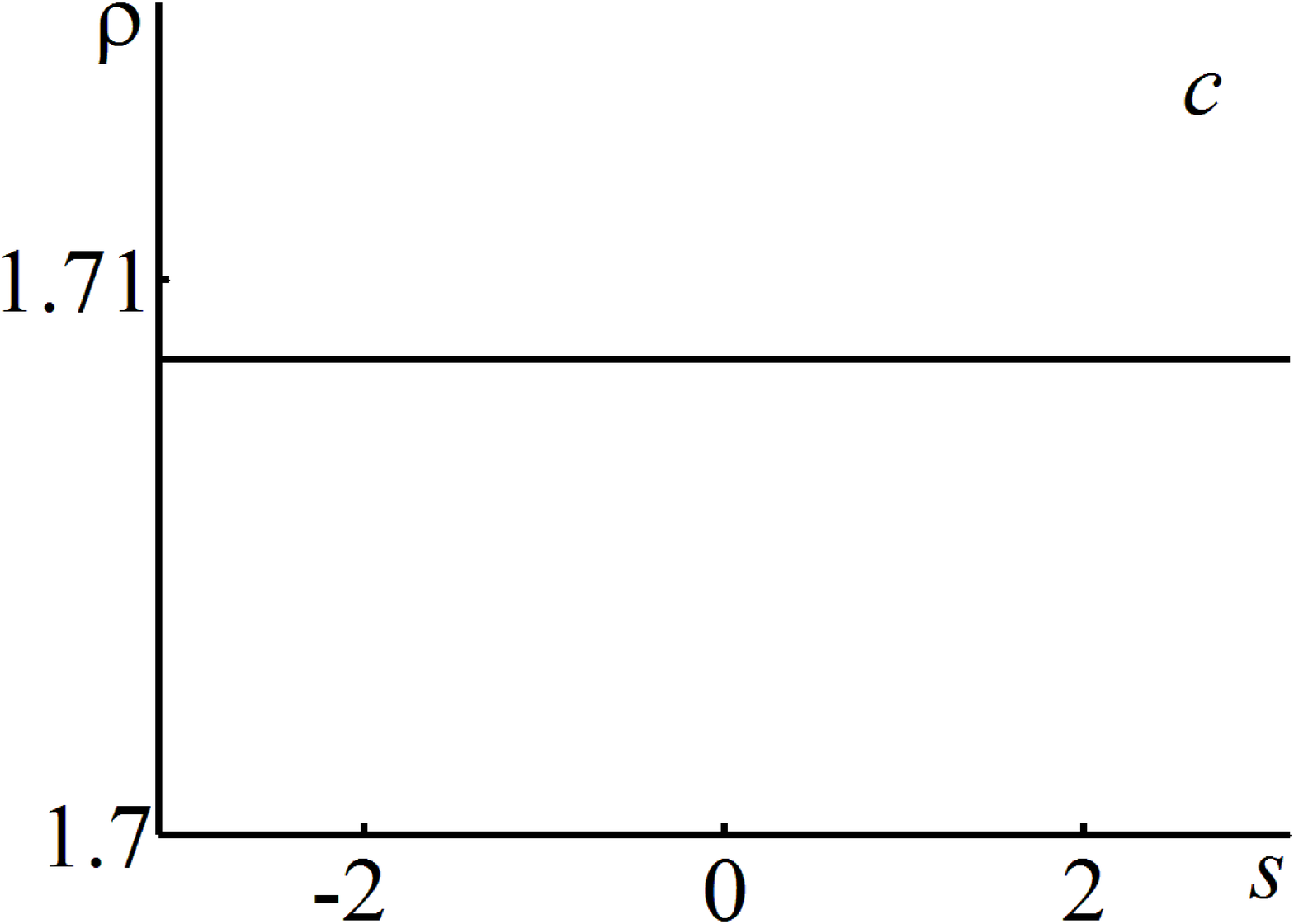}
\caption{Graphs of the function $\rho(t,s)$ for $t=100$,
$\gamma = 1$, and $D=0$ ($a$), 0.005 ($b$), and 0.5 ($c$)}
\label{dif}
\end{figure}

To show the part played by diffusion in pattern formation, we set an
initial condition of another type by using a cut-off function.
Figure~\ref{dif} displays the solution of equation \eqref{1d_1} obtained
by numerical simulation that is concentrated on a circumference for
$a=1$, $b_0=1$, $\varkappa=0.2$, $R=1$, $\gamma=1$, and
$\rho_\varphi(s) = \theta(s+2)-\theta(s-2)$, as a function of the
parameter $D$.

If the initial distribution is a cut-off function and $D=0$,
patterns are formed in a closed region where the population was
initially located (Fig.~\ref{dif}$a$). If $D$ is small enough,
patterns are formed but they experience qualitative changes because
of the expansion of concentration area (Fig.~\ref{dif}$b$). For large
values of $D$, pattern formation is not observed
(Fig.~\ref{dif}$c$).

\section{Conclusion}

The phenomenon of pattern formation in one-species population dynamics was studied
in many models based on the generalized FKPP equation with nonlocal interaction effects.
In this paper, we focus on the pattern formation of special type when
the patterns are concentrated on an evolving lower dimensional manifold $\Lambda_k^t$
whose dimension $k$ is less then the space dimension $n$.

This property allows us to describe such pattern formation in a more simple way.
Using ideas of the semiclassical approximation method for the nonlocal FKPP equation,
we characterize dynamics of formation of the concentrated patterns by
the Einstein-Ehrenfest system describing evolution of the manifold
$\Lambda_k^t$ and the semiclassically limited distribution $\rho(t,s)$.

We have found an exact solution of the dynamic equation determining the SLD $\rho(t,s)$.
This solution is space homogeneous and monotonically depends on time. By analogy with
\cite{naumkin1994,shismarev1999,komarov2001,komarov2002,komarov2011}, we assume
that the patterns above can be described as large-time perturbations of this exact solution.
The large-time asymptotics for the SLD $\rho(t,s)$ are constructed explicitly in the class
of functions which tend to the above exact solution as $T\to\infty$, to within $O(1/T^2)$.
Thereby, the exact solution can be regarded as an attractor of the constructed class of
the asymptotic solutions and, consequently, of the correspondent concentrated patterns.
Due to the patters evolve monotonically without qualitative changes to some steady-state,
we conclude that these asymptotic solutions describe approximately the quasi-steady-state patterns.
Note that the considered structures occur only under special choice of the model parameters.
The role of diffusion in structure formation is discussed.

The approach used in this paper allows one, on the one hand, to gain information on the
most essential characteristics of patterns and, on the other hand, to apply the methods
developed for one-dimensional problems to multidimensional problems. It should be
noted that the natural extension of the work is the problem of finding solutions $u(\vec x,t)$
to the FKPP equation using the functions $\vec X(t,s)$ and $\rho(t,s)$. The WKB-Maslov method
\cite{Maslov1,Maslov2,BelDob} provides conceptual way of finding an asymptotical solution
to the problem. Also, a direct study of Einstein-Ehrenfest system for $k>1$ is of interest.

The formalism proposed can be generalized to concentration manifolds of more general
topological structure, such as multiply connected manifolds \cite{Maslov1},
and to curved manifolds describing the growth of microbial populations on
complex structure objects.

\appendix
\def\thesection{Appendix~\Alph{section}}
\def\theequation{\Alph{section}.\arabic{equation}}
\section{}

Here, we show that substituting \eqref{vst_beta1} in (\ref{anal4y})
in view of \eqref{anal_resh19aa} and \eqref{vst_beta6}, we obtain
(\ref{vst_beta8}).  The coefficient of $v_j(s)$ in (\ref{vst_beta8})
is $\beta^{(1)}_j(t)$. To compare the solutions obtained by
\eqref{anal_resh3d} and \eqref{evolmn-dddd}, we calculate the
coefficient of $v_j(s)$ using (\ref{anal4y}) for the initial
distribution $\rho_\varphi(s)$ (\ref{vst_beta0}). The function
$\rho(t,s)$ is
\begin{align}
\rho(t,s)&=\rho_\varphi(s)\exp\bigg[at-\varkappa\lambda_0
v_0\int\limits_{0}^{t}\beta_{0}^{(0)}(t'){\rmd}t'\bigg] \times \nonumber\\
& \times \exp\bigg[-
\dfrac{1}{T}\sum_{l=-\infty}^{\infty}\varkappa\lambda_l
v_l(s)\int\limits_{0}^{t}\beta^{(1)}_{l}(t'){\rmd}t'\bigg]= \nonumber\\
& = \bigg(\beta_{00}v_0 + \dfrac 1 T
\sum_{l=-\infty}^{\infty}\beta_{1l} v_l(s)
\bigg)\exp\bigg[at-\varkappa\lambda_0
v_0\int\limits_{0}^{t}\beta_{0}^{(0)}(t'){\rmd}t'\bigg] \times \nonumber\\
& \times
\exp\bigg[-\dfrac{1}{T}\sum_{l=-\infty}^{\infty}\varkappa\lambda_l
v_l(s)\int\limits_{0}^{t}\beta^{(1)}_{l}(t'){\rmd}t'\bigg] = \nonumber\\
& = \bigg(\beta_{00}v_0 + \dfrac 1 T
\sum_{l=-\infty}^{\infty}\beta_{1l} v_l(s)
\bigg)\dfrac{\beta_{0}^{(0)}(t)}{\beta_{00}}
\bigg[1-\dfrac{1}{T}\sum_{l=-\infty}^{\infty}\varkappa\lambda_l
v_l(s)\int_{0}^{t}\beta^{(1)}_{l}(t'){\rmd}t'\bigg].
\end{align}

The coefficient of $v_j(s)$ is
\begin{align}
& \dfrac{\beta_{0}^{(0)}(t)}{\beta_{00}} \bigg(-\beta_{00}v_0
\varkappa\lambda_j \int\limits_{0}^{t}\beta^{(1)}_{j}(t'){\rmd}t'
+\beta_{1j}\bigg) = \nonumber\\
&\qquad = \dfrac{\beta_{0}^{(0)}(t)}{\beta_{00}}
\bigg(-\beta_{00}v_0 \varkappa\lambda_j
\int\limits_{0}^{t}\dfrac{\beta_{1j}{\rme}^{at'}}
{[1+\varkappa\lambda_0\beta_{00}(a\sqrt{2\pi})^{-1}
({\rme}^{at'}-1)]^{1+\lambda_j/\lambda_0}}{\rmd}t'
+ \beta_{1j}\Bigg) = \nonumber\\
&\quad = \dfrac{\beta_{0}^{(0)}(t)}{\beta_{00}} \bigg(\beta_{00}v_0
\varkappa\lambda_j\dfrac{\beta_{1j}}{\varkappa\lambda_0\beta_{00}v_0}
\dfrac{\lambda_0}{\lambda_j}\dfrac{1}{[1+\varkappa\lambda_0\beta_{00}
(a\sqrt{2\pi})^{-1}({\rme}^{at'}-1)]^{\lambda_j/\lambda_0}}\Bigg|_0^t+\beta_{1j}\Bigg)
= \nonumber\\
&\qquad = \dfrac{\beta_{0}^{(0)}(t)}{\beta_{00}}\beta_{1j}
\bigg(\dfrac{1}{[1+\varkappa\lambda_0\beta_{00}(a\sqrt{2\pi})^{-1}
({\rme}^{at}-1)]^{\lambda_j/\lambda_0}}-1+1\Bigg)= \nonumber\\
&\qquad
=\dfrac{\beta_{00}{\rme}^{at}}{[1+\varkappa\lambda_0\beta_{00}
(a\sqrt{2\pi})^{-1}({\rme}^{at}-1)]}\dfrac{\beta_{1j}}{\beta_{00}}
\dfrac{1}{[1+\varkappa\lambda_0\beta_{00}(a\sqrt{2\pi})^{-1}
({\rme}^{at}-1)]^{\lambda_j/\lambda_0}}= \nonumber\\
&\qquad = \dfrac{\beta_{1j}{\rme}^{at}}
{[1+\varkappa\lambda_0\beta_{00}(a\sqrt{2\pi})^{-1}({\rme}^{at}-1)]^{\lambda_j/\lambda_0+1}}
= \beta^{(1)}_j(t).
\end{align}

As the coefficient of $v_j(s)$ is also equal to $\beta^{(1)}_j(t)$,
we can state that the approach based on relations
(\ref{anal_resh3d}) is equivalent, to within $O(1/T^2)$, to that
based on relation \eqref{evolmn-dddd}.

\section{}

We show that the expansion of the function
$\rho^{(1)}(\theta,\tau,s)$ in the series (\ref{anal_resh21}) for
equation (\ref{anal_resh18b}) also yields (\ref{vst_beta8}). We
substitute (\ref{anal_resh21}) in equation (\ref{anal_resh18b}) and
obtain
\begin{multline}
a\sum_{l=-\infty}^{\infty}(C_l)_\theta v_l(s) -
a\sum_{l=-\infty}^{\infty}C_l v_l(s) + \varkappa\lambda_0v_0\beta_0^{(0)}(\theta)
\sum_{l=-\infty}^{\infty}C_l v_l(s) +\\
+\varkappa
v_0\beta_0^{(0)}(\theta)\sum_{l=-\infty}^{\infty}C_l\lambda_l v_l(s)
= - (v_0\beta_0^{(0)}(\theta))_\tau
\label{anal_resh22}
\end{multline}
where $\beta_0^{(0)}(\theta)$ is given by (\ref{anal_resh19aa}).

Multiplying (\ref{anal_resh22}) by $v_{-j}(s)$ and integrating it from $-\pi$ to $\pi$, we get
\begin{equation}
\begin{aligned}
& j=0, \quad a (C_0)_\theta - a C_0 + \varkappa\lambda_0v_0\beta_0^{(0)}C_0 + \varkappa \lambda_0 v_0\beta_0^{(0)} C_0 = 0,\\
& j\neq 0, \quad a (C_j)_\theta - a C_j +
\varkappa\lambda_0v_0\beta_0^{(0)} C_j + \varkappa \lambda_j v_0\beta_0^{(0)} C_j=0.\end{aligned} \label{anal_resh24}
\end{equation}

The solutions of equations (\ref{anal_resh24}) are
\begin{align}
j=0,\; & C_0 = C_0(0)\exp\bigg[\theta - 2\ln
\bigg(1+\dfrac{\varkappa\lambda_0v_0\beta_{00}}{a}({\rme}^{\theta}-1)\bigg)\bigg]
= \nonumber\\
& =
\dfrac{C_0(0){\rme}^\theta}{[1+\varkappa\lambda_0v_0\beta_{00}a^{-1}
({\rme}^{\theta}-1)]^2}
=\dfrac{C_0(0){\rme}^{at}}{[1+\varkappa\lambda_0v_0\beta_{00}a^{-1}
({\rme}^{at}-1)]^2},
\nonumber\\
j\neq0,\; & C_j = C_j(0)\exp\bigg[\theta -
\dfrac{\lambda_0+\lambda_j}{\lambda_0}\ln
\bigg(1+\dfrac{\varkappa\lambda_0v_0\beta_{00}}{a}({\rme}^{\theta}-1)\bigg)\bigg]
= \nonumber\\
&
=\dfrac{C_j(0){\rme}^\theta}{[1+\varkappa\lambda_0v_0\beta_{00}a^{-1}
({\rme}^{\theta}-1)]^{1+I_j(\mu)/I_0(\mu)}} = \nonumber\\
& =
\dfrac{C_j(0){\rme}^{at}}{[1+\varkappa\lambda_0v_0\beta_{00}a^{-1}
({\rme}^{at}-1)]^{1+I_j(\mu)/I_0(\mu)}}.\label{anal_resh26}
\end{align}

Thus, the final solution of equation (\ref{anal2}), to within $O(1/T^2)$,
can be written as
\begin{align}
\rho (\theta,\tau,s)&=
\dfrac{1}{\sqrt{2\pi}}\dfrac{\beta_{00} {\rme}^\theta}{1+\varkappa\lambda_0v_0\beta_{00}a^{-1}
({\rme}^\theta-1)}+ \nonumber\\
& + \dfrac{1}{T\sqrt{2\pi}}
\sum_{l=-\infty}^{\infty}\dfrac{C_l(0){\rme}^\theta
{\rme}^{ils}}{[1+\varkappa\lambda_0v_0\beta_{00}a^{-1}
({\rme}^{\theta}-1)]^{1+I_l(\mu)/I_0(\mu)}}+\ldots = \nonumber\\
& = \dfrac{1}{\sqrt{2\pi}}\dfrac{\beta_{00}{\rme}^{at}}
{1+\varkappa\lambda_0v_0\beta_{00}a^{-1}({\rme}^{at}-1)}+ \nonumber\\
& +\dfrac{1}{T\sqrt{2\pi}}
\sum_{l=-\infty}^{\infty}\dfrac{C_l(0){\rme}^{at}
{\rme}^{ils}}{[1+\varkappa\lambda_0v_0\beta_{00}a^{-1}
({\rme}^{at}-1)]^{1+I_l(\mu)/I_0(\mu)}}+\ldots, \label{anal_resh27b}
\end{align}
which is the same as \eqref{vst_beta8} for $C_l(0)=\beta_{1l}$.

\setcounter{section}{1}

\section*{Acknowledgments}

The work was partially supported by the Russian Federal program ''Kadry'' under contracts
16.740.11.0469 and program 'Nauka' No 1.604.2011.

\end{document}